\documentclass[a4paper,11pt]{article}
\pdfoutput=1 

\usepackage{jheppub} 

\usepackage{graphicx}
\usepackage[figurename=Fig.]{caption}
\usepackage[tablename=Tab.]{caption}
\usepackage{color}
\usepackage{float}
\usepackage{amssymb}
\usepackage{amsmath}
\usepackage{mathrsfs}
\usepackage{textcomp}
\usepackage{bbm}
\usepackage{cases}
\usepackage{makecell}
\usepackage{pdflscape}
\usepackage{subfigure}
\usepackage{pdfpages}
\usepackage{makeidx}
\usepackage{bm}
\usepackage{xstring}
\usepackage{lipsum}
\usepackage{slashed}
\usepackage{multicol}
\usepackage{setspace}
\usepackage{booktabs}
\usepackage[table]{xcolor}
\usepackage{multirow}
\usepackage{braket}
\usepackage[separate-uncertainty]{siunitx}
\usepackage[displaymath,mathlines]{lineno}

\def\up{\mathrm}

\renewcommand{\thetable}{\Roman{table}}

\graphicspath{{PDF/}}

 \usepackage{etoolbox}          

\newcommand*\linenomathpatch[1]{%
  \cspreto{#1}{\linenomath}%
  \cspreto{#1*}{\linenomath}%
  \csappto{end#1}{\endlinenomath}%
  \csappto{end#1*}{\endlinenomath}%
}
\newcommand*\linenomathpatchAMS[1]{%
  \cspreto{#1}{\linenomathAMS}%
  \cspreto{#1*}{\linenomathAMS}%
  \csappto{end#1}{\endlinenomath}%
  \csappto{end#1*}{\endlinenomath}%
}

\expandafter\ifx\linenomath\linenomathWithnumbers
  \let\linenomathAMS\linenomathWithnumbers
  \patchcmd\linenomathAMS{\advance\postdisplaypenalty\linenopenalty}{}{}{}
\else
  \let\linenomathAMS\linenomathNonumbers
\fi

\linenomathpatch{equation}
\linenomathpatchAMS{gather}
\linenomathpatchAMS{multline}
\linenomathpatchAMS{align}
\linenomathpatchAMS{alignat}
\linenomathpatchAMS{flalign}

\makeatletter
\patchcmd{\mmeasure@}{\measuring@true}{
  \measuring@true
  \ifnum-\linenopenaltypar>\interdisplaylinepenalty
    \advance\interdisplaylinepenalty-\linenopenalty
  \fi
  }{}{}
\makeatother

\title{\boldmath Meson properties and symmetry emergence based on the deep neural network}

\author[a]{Xin Tong,}
\author[b]{Wei Feng,}
\author[c,d]{Weiwei Xu,}
\author[e,f]{Chao-Hsi Chang,}
\author[g]{Guo-Li Wang,}
\author[a]{Qiang Li\footnote{Corresponding author}}

\affiliation[a]{School of Physical Science and Technology, Northwestern Polytechnical University, Xi'an 710072, China}
\affiliation[b]{School of Information Mechanics and Sensing Engineering, Xidian University, Xi\rq{}an, 710071, China}
\affiliation[c]{Shandong Institute of Advanced Technology, 250100, Jinan, China}
\affiliation[d]{Shandong University, 250100, Jinan, China}
\affiliation[e]{CAS Key Laboratory of Theoretical Physics, Institute of Theoretical Physics, Chinese Academy of Sciences, Beijing 100190, China}
\affiliation[f]{School of Physical Sciences, University of Chinese Academy of Sciences, Beijing 100049, China}
\affiliation[g]{Department of Physics, Hebei University, Baoding 071002, China}

\emailAdd{liruo@nwpu.edu.cn}

\abstract{
As a key property of hadrons, the total width is quite difficult to obtain in theory due to the extreme complexity of the strong and electroweak interactions. In this work,  a deep neural network model with the Transformer architecture is built to precisely predict  meson widths in the range of $10^{-14} \sim 625$\,MeV based on meson quantum numbers and masses.  The relative errors of the predictions are ${0.12\%, 2.0\%,}$ and {$0.54\%$} in the training set, the test set, and all the data, respectively. We present the predicted meson width spectra for the currently discovered states and some theoretically predicted ones. The model is also used as a probe to study the quantum numbers and inner structures for some undetermined states including the exotic states. Notably, this data-driven model is investigated to spontaneously exhibit good charge conjugation symmetry and approximate isospin symmetry consistent with physical principles. The results indicate that the deep neural network can serve as an independent complementary research paradigm to describe and explore the hadron structures and the complicated  interactions in particle physics alongside the traditional experimental measurements, theoretical calculations, and lattice simulations. 

}


\begin{document} 

\maketitle
\flushbottom

\section{Introduction}

Particle physics aims to understand the essence of fundamental particles and their interactions in nature. Quarks and gluons are bound by strong interactions to form hadrons, namely, mesons and baryons. Due to quark confinement\,\cite{HanMY1965,Fritzsch1973},  hadron properties play key roles in understanding the strong interactions, which are directly determined by the dynamics of quarks and gluons inside. The strong interactions are described by the quantum chromodynamics\,(QCD). In the high-energy region (short distance), QCD's perturbation calculations have achieved great success, and its theoretical predictions are highly consistent with experimental results, which becomes one of the most important experimental verifications of the Standard Model for particle physics. However, in the low-energy region\,(long distance), the non-perturbative effects of QCD make analytical calculations quite difficult, which leads to making predictions of hadron lifetimes quite difficult in theory.

Mesons as the quark-antiquark bound states play important roles for studying the non-perturbative QCD interactions. The meson masses and decay widths are hence direct reflections of the internal interaction dynamics. Generally speaking, the total width of a meson is a summation of all the possible decay channels, while the decay behaviors are directly dependent on the meson quantum numbers and the involved interactions. Research on mass spectra and total widths provides rich information on the internal structure and interactions of mesons. 

There exist two main goals in hadron physics. One is to obtain the hadron mass and decay behaviors based on the quantum numbers, while the other is to obtain the quantum numbers from the observed mass and lifetime\,(or total width equivalently).  With the QCD-inspired potentials and quark model, by solving the corresponding bound states equation, mass spectra for  traditional hadrons can be well established in theory\,\cite{GI1985,Capstick1986,Eichten1994} though some exotic hadrons discovered in recent years are not included\,\cite{Klempt2007,ChenHX2016,Olsen2017,GuoFK2018,Karliner2018}. However,  it is quite difficult to obtain the total widths based on the traditional calculations of Standard Model dynamics for the extreme theoretical complications.  Hence it is quite necessary and helpful  to predict lifetime spectra from a new reliable method.

In recent years, with the continuous operation of high-precision experiments such as LHCb, CMS, ATLAS, Belle II, and BESIII, significant progress has been achieved in hadron spectroscopy research, and a large number of new hadron states have been discovered including quite a lot of exotic state candidates\,\cite{PDG2024}. These new discoveries greatly enrich our understanding of the hadron family, but at the same time pose a serious challenge to the current theoretical framework. The internal structures, quantum numbers, masses, and decay widths of many particles have not yet been fully determined\,\cite{Klempt2007,ChenHX2016,Olsen2017,GuoFK2018,Karliner2018,ChenHX2023,PDG2024}. Therefore, systematic research on the meson spectra, especially reliable predictions for the total widths has become quite an important problem in hadron physics. All these challenges mentioned above highlight the necessity of exploring alternative data-driven methods.

On the other hand, deep learning allows computational models that are composed of multiple processing layers to learn representations of data with multiple levels of abstraction\,\cite{LeCun2015}. 
In the last ten years the deep neural networks have become a powerful paradigm for many complex problems since several key breakthroughs happened in this field. The application of back-propagation algorithm laid the theoretical foundation for effectively training neural networks\,\cite{Rumelhart1986}. Then a series of powerful architectures have made great achievements in computer science areas, such as the recurrent neural network\,(RNN)\,\cite{Hopfield1982,Elman1990}, the long short-term memory network\,(LSTM)\,\cite{Hochreiter1997} and neural probabilistic representation\,\cite{Bengio2003} for natural language and other sequential data processing, and the convolutional neural network\,(CNN)\,\cite{LeCun2002} for image recognition. The residual networks\,(ResNet) overcame the bottleneck of gradient vanishing in deep networks, making it more effective to construct and train extremely deep networks\,\cite{HeKM2016}. This series of advancements ultimately gave birth to the Transformer architecture, which is considered as the most advanced and powerful  architecture characterized by its self-attention mechanism\,\cite{Vaswani2017}. 


Currently the deep neural networks have also become a powerful paradigm for solving complex problems in high energy physics\,\cite{Krenn2022,Radovic2018,Guest2018,Carleo2019,ZhouK2023,Ma:2023pts}. Its core capability lies in extracting physical information from high-dimensional complex experimental data, which is applied in particle track reconstruction\,\cite{ExaTrkX2020}, jets classification\,\cite{Lonnblad1990,Bols2020,Komiske2019}, the search for new physics\,\cite{Roe2005,Baldi2016}, and mass spectra of atomic nuclei\,and hadrons\,\cite{Mumpower2022,HeW2023,zhang:2025mlu,Zhang:2022gmo}, etc. Other data-driven paradigms have also shown great potential. The symbolic regression has been applied to explicitly determine the global behavior of the QCD strong coupling constant\,\cite{Wang2024CPL}, and physics-informed neural networks have been employed to analyze chromoelectric flux tubes\,\cite{Kou2025PRD}. These works, together with deep learning approaches for hadron spectroscopy, illustrate that machine learning is becoming a versatile tool for addressing diverse problems in strong interaction physics.

Encouraged by these advances, several recent researches have also begun to directly focus on meson width topic. A deep neural network (DNN) model has been developed to predict mass and width of meson states\,\cite{Malekhosseini2024}; another work compared the performance of various machine learning algorithms on this problem\,\cite{Akan2024}. These studies have validated the feasibility of data-driven approaches on this issue, but they also face several core challenges that constrain their development, for example, the performance of deep learning models is highly dependent on  accurate hadron encoding and large-scale datasets,  while the experimental hadron data is relatively sparse and there also exists a large number of degeneracy.
This work aims to systematically explore a data enhancement strategy based on physical principles, and examine whether a deep neural network model trained in this way can successfully predict the meson widths and further reveal the underlying physical symmetries behind the data. To this end, we have developed a deep learning framework that starts with a systematic feature engineering approach to address meson encoding. The framework adopts the Feature Tokenizer Transformer\,(FT-Transformer) architecture\,\cite{Gorishniy2021} as the core model and enhances the data by the Gaussian Monte-Carlo method to effectively train the model and improve its robustness.  

The trained model demonstrates strong predictive ability within a huge range spanning nearly 17 orders of magnitude, from about $10^{-14}$\,MeV to about 625 MeV. The relative error on an unprecedented conventional meson test set reaches the level of one percent, demonstrating excellent generalization performance. And we further verified the physical consistency of the model. Our research shows that this data-driven model can successfully reproduce the fundamental symmetry in particle physics without any prior physical constraints. The predicted results not only strictly follow the charge conjugation symmetry, but also quantitatively reflect the approximate  isospin symmetry for meson widths.

In addition, the model exhibits different response to particles with different internal structures, which also provides us a new insight to explore hadrons. We find that although the model is able to learn the patterns of traditional mesons and some exotic states contained in the training data, its predictions exhibit significant biases to the well-known exotic state candidates such as $D_{s0}^{*} (2317)$ and $\chi_{c1} (3872)$. It then suggests that these states may have different internal structures than the particles in the training set, which also suggests that the model's predictions may serve as a data-driven probe to assist in identifying and classifying potential non-traditional hadron states.

This paper is organized as follows. In Section\,\ref{sec-2}, we first state the used data, data enhancement method, and the data encoding scheme. Then in Section\,\ref{sec-3} we introduce the neural network model we used to deal with the meson width problems, the loss function, the train methods and also the hyper parameters. The obtained results, predictions, and discussions are presented in Section\,\ref{sec-4}. Finally we give a brief summary and outlook.

\section{Meson encoding methods and data enhancement}\label{sec-2}

In a prediction task based on a deep neural network, the performance of the model is highly dependent on the quality of the input data and representation of the data features. In this section, we introduce the complete procedure of data processing for the meson width predictions. First, we construct a complete feature vector which can definitely and precisely represent quantum states of mesons; and then, we explore a data enhancement strategy to solve the important issue of limited data for deep learning.

\subsection{Features used to denote the mesons}

A particle is represented by a set of quantum numbers, such as spin $J$, parity $P$, charge conjugation $C$, $G$, isospin $I$ and its third component $I_3$, the quark contents, mass and total width\,(or lifetime equivalently). These quantum numbers will also be called features within the machine learning terminology.  The non-relativistic quantum numbers, inner spin $S$ and orbital angular momentum  $L$, are often used to denote the mesons in literature, however, these two are neither good quantum numbers in describing the particles nor the physical observables in experiments. We will not use the non-relativistic $S$ and $L$ as the meson features to train the model. 

According to the conventions in deep learning, we categorize the aforementioned features into two groups. One group describing the inner symmetry of the mesons consists of  the categorical features, including the $P$, $C$ and $G$; and the other group consists of the numerical features, including the $I$, $I_3$, $J$, and the quark flavor coefficients. The features in the former one are assumed to be numerical-independent and different values just indicate different categories, while the latter ones are assumed to be numerical-dependent. It should be pointed out that the above category scheme is not rigorous. 

{Also, for some mesons, the information of the experimental data is not complete while some quantum numbers may be missing. We apply the following scheme to deal with the case of missing features. For the categorical features, the inapplicable features and the undetermined features are considered as two different independent discrete features respectively, for example, the values of $C$-parity for $\pi^+$\,(inapplicable) and $X(4160)$\,(undetermined currently) are assigned as two different categorical labels. For the numerical features, we developed a dual-path embedding framework that decouples a feature's value from its certainty. We first transform the original numerical value $x_\up{num}$ into a number-mask pair as
\begin{equation}
(x'_\up{num}, x_\up{mask}) =
\left\{
\begin{aligned}
&(0,0) & & ~~~ \text{undetermined }x_\up{num} , \\
&(x_\up{num},1)&& ~~~\text{otherwise}.
\end{aligned}
\right.
\end{equation}  
Then every numerical feature is coupled with an independent feature which gives the undetermined data a specific mask value 0 to distinguish it from the usual sample. This pair is then processed by an augmented embedding module. Namely, the embedding of this number-mask pair will then be fed into a linear layer to give a $d$-dimensional fused feature vector which encodes both the value and its certainty.}

\subsubsection{Quark contents encoding}
The first challenge in using neural networks to deal with the hadron spectroscopy is the encoding of quark contents inside hadrons. Within the traditional quark model, mesons consist of a quark-antiquark pair. A direct encoding scheme is to use a 10-dimensional binary vector to represent the five quarks and the corresponding antiquarks, namely, quark contents can be expressed as 
\begin{gather}
\bm{q}= (u,\bar u, d, \bar d, s, \bar s, c, \bar c, b, \bar b),
\end{gather}
where the $i$th element $q_i=1$  denotes that the corresponding quark exists in the meson while $q_i=0$ not;  we did not take into account the (anti)top quark. However, this naive binary encoding cannot describe the mesons in flavor superposition states. The quark content of many neutral mesons is a superposition of several quark-antiquark pairs, for example, the flavor wave function of neutral pion is usually expressed as
$\frac{1}{\sqrt{2}} (u\bar u - d\bar d)$.
In order to represent the information of these flavor superposition states, we directly pass the corresponding flavor coefficients into the quark content vector $\bm{q}$, for example, the $\pi^0$ and $\eta$ are encoded as
\begin{gather}
\bm{q}_{\pi^0} = \textstyle(\frac{1}{\sqrt2}, \frac{1}{\sqrt2},\frac{1}{\sqrt2},\frac{1}{\sqrt2},0,0,0,0,0,0 ), \\
\bm{q}_{\eta} = \textstyle(\frac{1}{\sqrt6}, \frac{1}{\sqrt6},\frac{1}{\sqrt6},\frac{1}{\sqrt6},\frac{2}{\sqrt6},\frac{2}{\sqrt6},0,0,0,0 ).
\end{gather}
Notice that here we only care about the absolute values of the coefficients which are aimed to reflect the weights of the corresponding quarks inside the mesons. 
It is important to clarify that the physical information carried by the relative phase (such as the distinction between $\rho^0$ and $\omega$) is not lost within the current flavor encoding scheme; instead, it is intrinsically related to the input quantum numbers $I$ and $I_3$, for example, $I_\rho=1$ and $I_\omega$=0. For complex mixed states where phase assignment is ambiguous\,(e.g., $\eta$-$\eta'$ mixing), the model relies on the mass features and other quantum numbers for differentiation. This encoding scheme ensures that the model receives sufficient discriminative information while maintaining numerical stability. However, it should be pointed out the fact that the quark coefficients are not good quantum numbers in experiments. The PDG only gives a universal flavor representation $[c_1(u\bar u+d\bar d)+c_2s\bar s]$ for the light unflavored mesons with $I=0$. It is worth further studying a more effective and accurate encoding scheme for the quark contents.

This encoding scheme for the quark contents can simplify the structure of the input features, and then the task for grasping deep symmetry information of the mesons can then be finished by the deep neural network with powerful self-attention mechanism.  
The encoding scheme can be naturally extensible to multiquark states. For instance, the fully charmed tetraquark candidate $T_{c\bar{c}1}(3900)^+$ is represented by
\begin{gather}
	\bm{q}_{T_{c\bar{c}1}(3900)^+} =\textstyle  (1, 0, 0, 1, 0, 0, 1, 1, 0, 0),
\end{gather}
where the four non-zero entries correspond to the $u$, $\bar d$, $c$ and $\bar c$ quarks, respectively. The encoding scheme can uniformly process conventional mesons and exotic states within the same feature space.
\subsubsection{Complete input features}

Another challenge  is that the aforementioned quantum numbers cannot definitely and uniquely represent a particle state. Several hadrons may share the same quantum numbers $(J,P,C,G,I,I_3)$ and quark contents but correspond to different masses and widths. There is degeneracy if we just use the features above. For example, $\rho(770)$ and $\rho(1450)$ share exactly the same quantum numbers $I^G(J^{PC})=1^+(1^{--})$ and quark contents $[u\bar d]$ but have quite different masses and widths. If  these degenerate features were used as input, it would lead to the one-to-many ambiguous mappings and lower the performance of the model. 

Within the quark model, these ambiguities may originate from several aspects. On the one hand, different principal quantum numbers cause different particle masses, such as $J/\psi$, $\psi(2S)$, $\psi(3770)$, $\psi(4040)$, $\psi(4160)$, $\psi(4260)$, $\psi(4360)$, and $\psi(4415)$, $\psi(4660)$. The $\psi$ family mesons all share the same $J^{PC}=1^{--}$ and quark content $[c\bar c]$. Although the usual principal quantum number $n$ may be used to label different states, there are several reasons prevent us using this feature. First, $n$ is a non-relativistic, model-dependent quantity;  second, $n$ is not a directly experimental observable; and finally, it is difficult to precisely establish all the principal quantum numbers of these mesons, since there may also exist  the so-called $2S$-$1D$, $3S$-$2D$, ..., mixing effects, namely, even the non-good principal quantum number $n$ is missing for most mesons.
On the other hand, the unnatural parity $(J^P=1^+,2^-,3^+,\cdots)$ mesons also have the extra inner freedom which makes the corresponding masses undetermined from the current features, for example, the $J^P=1^+$ charmed mesons $D_1(2420)$ and $D_1(2430)$, and the charmed-strange meoson $D_{s1}(2536)$ and $D_{s1}(2460)$. These mesons are usually considered as the two physical states from the non-relativistic $^1L_L$-$^3L_L$ mixing, and are distinguished by an extra introduced  mixing angle parameter. However, this parameter is  also a model-dependent quantity and cannot be directly measured  from experiments. Hence a definite quantum number is also missing to represent the inner freedom of unnatural parity mesons, which makes it impossible to determine the masses and widths together for such mesons from a set of good quantum numbers.

All these ambiguities make it difficult to directly use the aforementioned features to train a deep neural network model to predict the meson masses and widths. On the other hand, the meson mass spectra can be well obtained by the quark model calculations\,\cite{GI1985}, and a more urgent and difficult problem is to calculate the meson widths. Then the mass itself can behave as a powerful input feature and be used to solve the  degenerate issues above. Compared to the total width, the meson masses are much easier to obtain based on the quark model calculations or the lattice QCD simulations, which has also become part of the motivation to predict the meson widths while not the masses. Notice here we do not distinguish the molecular states or the compact tetraquark ones for the exotic states with the same definite quantum numbers. The different interpretation pictures are model-dependent and are not experimental observable good quantum numbers, and hence will not be used as the input features for neural network. All the differences of widths for those states are attributed to the mass differences.

Additionally, we also introduce an auxiliary feature $N$ with continuous values to represent the mass ordering, which is defined by normalizing the central value of the meson mass with $m_{\pi^0}$. Then the feature $N$ behaves as an extensible relative one and any new discovered meson can be easily embedded into this system. Notice that $N$ is not an independent feature but an auxiliary feature derived from the meson mass, designed to help the model better interpret the mass spectrum. Finally, we can encode a meson by a feature vector as
\begin{gather}
\bm{v} = (J,P,C,G,I,I_3,u,\bar u, d, \bar d, s, \bar s, c, \bar c, b, \bar b,m,N).
\end{gather}
And our aim is to use the above feature vector $\bm{v}$ to predict the meson total width $\Gamma$ based on the deep neural network.

The theoretical feasibility of applying deep neural networks to predict meson widths is supported by the Universal Approximation Theorem\,\cite{Hornik:1989yye,Cybenko:1989iql,Lu:2017yye}, which says that a deep neural network with sufficient complexity can approximate any function to arbitrary accuracy. Physically, the meson width is uniquely determined by its mass and the complete set of quantum numbers, which encode the selection rules governing possible decay channels. The representations of different particles in feature space are fundamentally distinct, which then provide necessary discriminative information for the neural network. Here we take the $\pi^0$ and $\pi^\pm$ as an example to show the relation between the feature representation and the meson quantum numbers. Specifically, $\pi^0$ is encoded with $I_3=0$, the definite $C$-parity\,($C=+1$), and the mixed quark flavor ${\pi^0}=\frac{1}{\sqrt{2}}(u\bar{u}-d\bar{d})$; in contrast, $\pi^\pm$ are encoded with $I_3=\pm 1$, a distinct flavor $(u\bar{d})/(d\bar{u})$, and a categorical embedding indicating that $C$-parity is not applicable. The definite $C$-parity permits electromagnetic decay while charged light mesons are restricted to weak decays. The neural network can then learn this non-linear mapping from the training data to predict the width property. The differences in the input quantum numbers  essentially define and determine the three states. These features are fully sufficient to distinguish the three states. On the other hand, the Universal Approximation Theorem guarantees that a sufficiently deep neural network can approximate this complex mapping, enabling the deep neural network to predict the meson widths bypassing the difficulties of direct non-perturbative QCD calculations.

\subsection{Gaussian Monte Carlo data enhancement}

The data used for training and testing in this work are from the experimental meson data in PDG 2024 and 2025 update\,\cite{PDG2024}, which include 400 mesons. The most important challenge for this task is the lack of data. Though more than 400 mesons have been discovered in experiments, the dataset is still quite small for training a deep neural network with good generalization ability. To overcome the bottleneck of data sparsity, data augmentation is considered an effective strategy. There are two main implementation paths in machine learning: one is to use deep generative models to learn and generate new synthetic samples\,\cite{Malekhosseini2025}; the second is to generate physically reasonable pseudo-data by Monte-Carlo sampling of experimental data. The latter one can expand the limited datasets and also improve model robustness. Therefore, it is widely used in particle physics, nuclear physics, and astrophysics\,\cite{ATLAS2024,LHCb2013A,Fujimoto2021,LIGO2016,ATLAS2023}. These two paths provide complementary perspectives for addressing data sparsity issues.

To overcome this problem, we implement the Gaussian Monte-Carlo data enhancement based on the uncertainties of experimental data in this work. The experimental measurement of a meson mass $m$ is always accompanied by the experimental uncertainties $\Delta m$ from both statistical and systematic errors, which in turn define the confidence interval of the corresponding experimental data. The core idea of the Gaussian Monte-Carlo data enhancement is to produce the Monte Carlo\,(MC) mass data for every hadron based on the Gaussian distribution with the standard deviation $\sigma$ defined by the experimental error $\Delta m$. 

The experimental measurement of a hadron mass in PDG is usually denoted as $m=({M_c}^{+\Delta m_+}_{-\Delta m_-})$ with $M_c$ representing the central value while $\Delta m_\pm$ denote the asymmetry errors.
The concrete procedures are as follows. First, we naively average the asymmetry errors $\Delta m_+$ and $\Delta m_-$ to obtain the standard deviation $\sigma=\frac12(\Delta m_+ +\Delta m_-)$. Then we assume the mass data follow the Gaussian distribution $\mathcal{N}(M_c,\sigma^2)$. Finally, for each meson, we implement the MC sampling to obtain $N_\up{aug}$ mass data within the range of $M_c\pm k\sigma$, where $k$ is the sampling constant used to control the distribution range of meson mass, and  $k=1$ and $N_\up{aug}=500$ are set in the current research.

\begin{figure}[h!]
\centering
\includegraphics[width = 0.490\textwidth, angle=0]{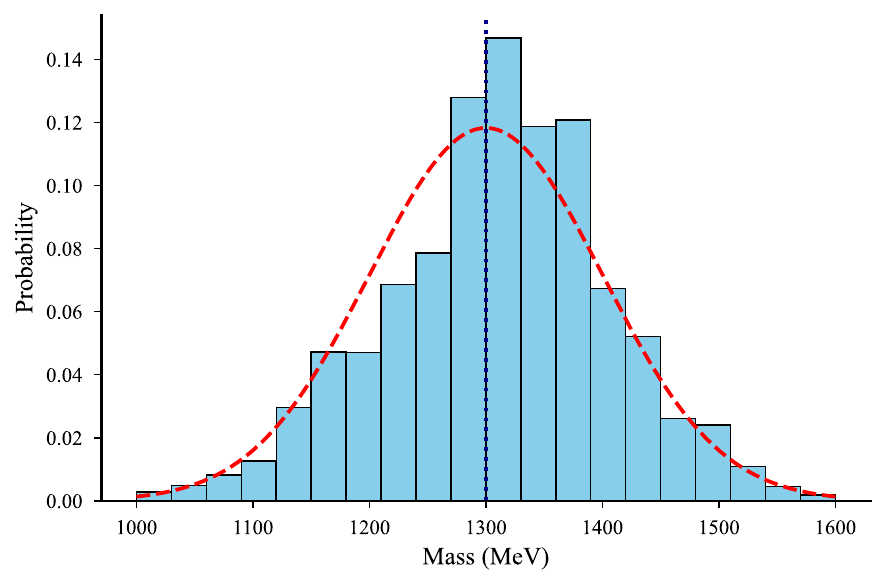}
\includegraphics[width = 0.490\textwidth, angle=0]{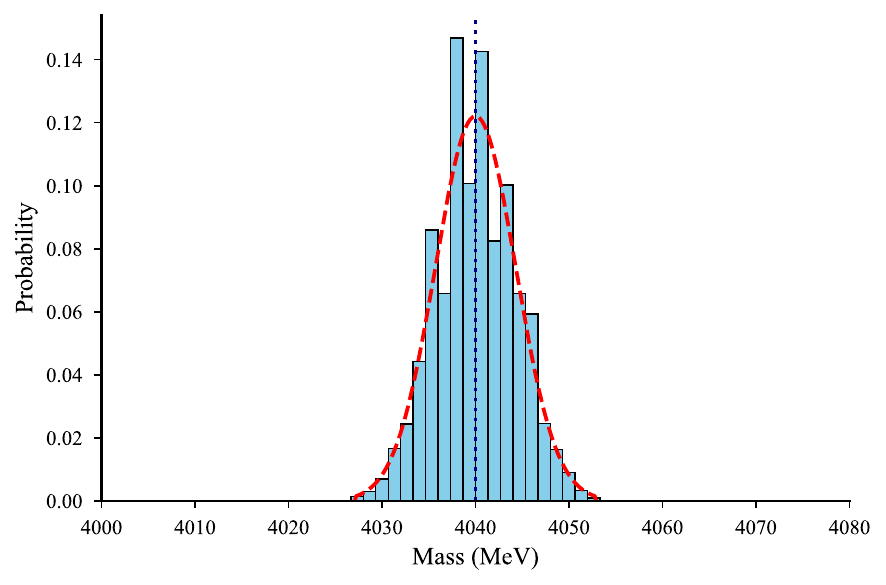}
\caption{The mass distribution of $\pi(1300)$ and $\psi(4040)$ after the Gaussian enhancement.}\label{F-Gauss} 
\end{figure}

All data points produced by MC sampling from an original meson share the same quantum numbers, with only the mass feature being varied. After this data enhancement, each meson state is effectively expanded, and the total data sample is increased from $\sim400$ to ${\sim200000}$. This process simulates the statistical nature of experimental detections and efficiently enhances the robustness of the neural network model. In \autoref{F-Gauss} we show the resulting mass distributions for $\pi(1300)$ and $\psi(4040)$. While the augmented mass values introduce a necessary variance, this can also obscure the precise hierarchy between different meson states. The auxiliary feature $N$ is therefore particularly important in this context, as it provides a stable, normalized index of each meson's position in the mass spectrum, complementing the fluctuating mass value. During the inference stage, the model utilizes the standard experimental central values\,(as listed in the PDG) as input without augmentation to produce deterministic width predictions.

\section{Deep neural network model} \label{sec-3}

After finishing the feature construction and data enhancement, we now choose a suitable neural network architecture and build the deep learning model to deal with the task for  meson width predictions. We will introduce the model architecture, the loss function, and the training and testing strategy.

\subsection{Backbone of the model}
\begin{figure}[h!]
\centering
\includegraphics[width = 1\textwidth, angle=0]{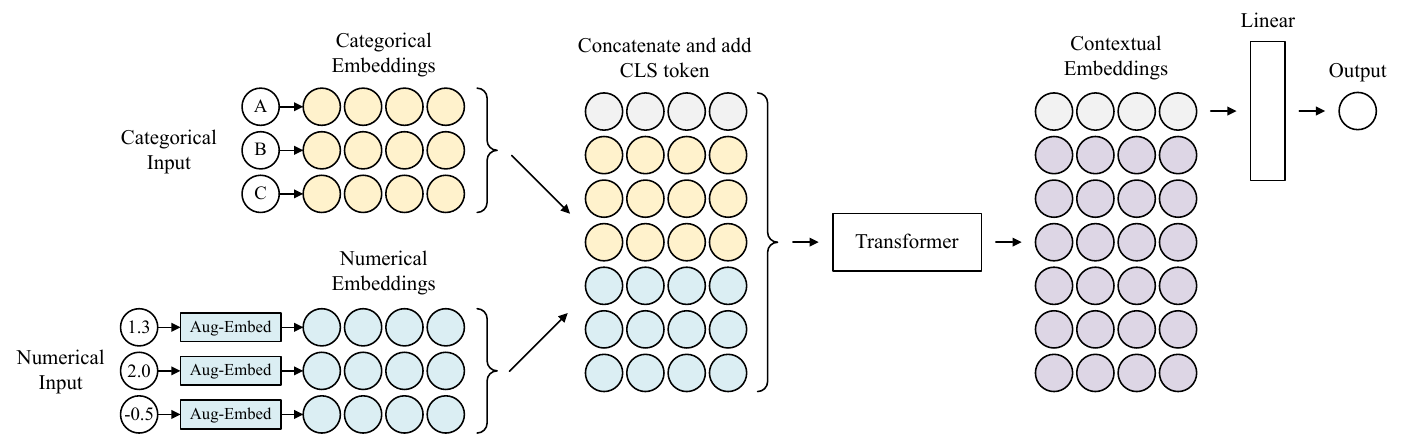}
\vspace{0.5em}
\caption{The main parts of the FT-Transformer architecture. Both the numerical and categorical features are first embedded to $d$-dimensional vectors, which are then together with the [CLS] token fed into the typical Transformer encoder layer.}\label{F-FT-Transformer} 
\end{figure}

Among all these deep neural network architectures, the Transformer\,\cite{Vaswani2017} model has outstanding performance in many aspects for its self-attention mechanism. Compared with the traditional fully connected network or the convolutional neural network, the Transformer model is able to process the input data in parallel and can also capture the long distance  dependencies of data. Though designed for the natural language processing, its core idea has also been successfully applied to many science fields to deal with the deeply complicated inner mapping and dynamics. Considering the fact that we are dealing with the tabular data with complicated heterogeneous features, we choose the Feature Tokenizer Transformer\,(FT-Transformer) as the backbone of the model. The FT-Transformer is designed specifically for tabular data and can effectively capture complex nonlinear dependence between different features by the self-attention mechanism. The main parts of the FT-Transformer are shown in \autoref{F-FT-Transformer}.

The FT-Transformer first makes feature tokenization to map the heterogeneous raw input features\,(both the categorical and continuous features) to a unified high-dimensional vector space.  For each categorical feature (such as particle quantum numbers $P$, $C$, and $G$), the model converts the discrete integer index into a dense feature vector $\bm{e}_\up{cat} \in \mathbbm{R}^d$ through an embedding layer with $d$ representing the embedding dimension. The embedding is implemented by a lookup table for the categorical features.

\begin{figure}[h!]
\centering
\includegraphics[width = 1.0\textwidth, angle=0]{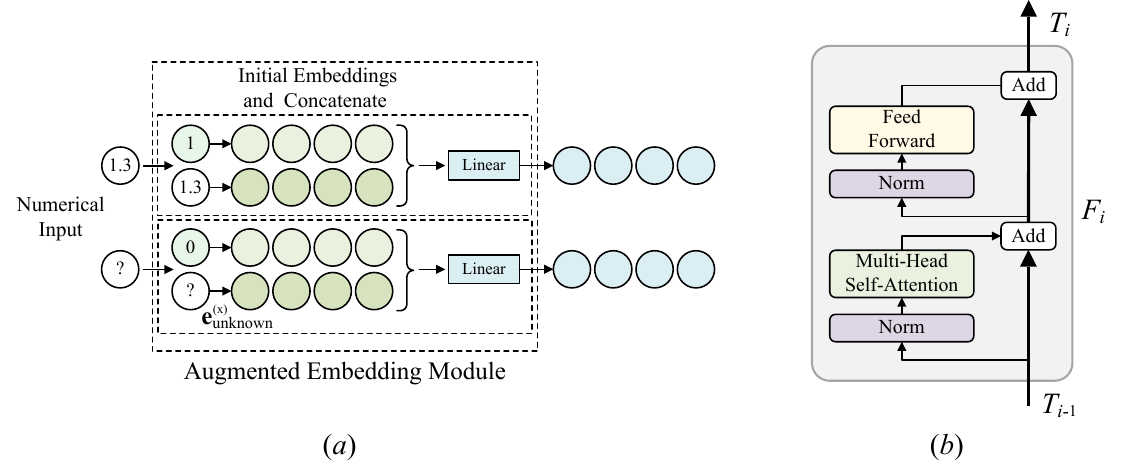}
\caption{(a) The Augmented Embedding module to deal with the numerical features; (b) one Transformer layer.}\label{F-one-Transformer} 
\end{figure}

{For each continuous feature\,(such as meson mass), the model employs an augmented embedding module detailed in Sec. 2.1 and illustrated in \autoref{F-one-Transformer}(a), which performs a linear transformation on each continuous feature value and then maps it into the same $d$-dimensional space, namely, 
\begin{gather}
\bm{e}_\up{num} =  {x}_\up{num} \bm{w}+ \bm{b},
\end{gather}
where $\bm{w},\bm{b}\in \mathbbm{R}^d$ denote the weight and bias vectors exclusive to the corresponding feature $x_\up{num}$. In addition, in order to enable the model to effectively utilize all the information of the raw data, we designed a mask value for each numerical feature to indicate whether this value is determined. Then we concatenate the embedding vectors of each feature-mask pair, and then fuse them through a linear layer to generate a single vector, namely, 
\begin{gather}
\bm{e}_{\up{fused}} = \text{Linear}[(\bm{e}_\up{raw}; \,\bm{e}_\up{mask})],
\end{gather}
where the symbol `;' in above bracket denotes the concatenation operation. The above process ensures that the model can deal with a feature with determined or undetermined value. }


All tokenized feature vectors will be concatenated with a special classification token [CLS] in the sequence dimension to form the final input sequence. The sequence is then fed into a Transformer encoder consisting of $L$ identical layers stacked together. Each layer of the encoder consists of two core submodules: a Multi-Head Self-Attention module and a feedforward network, which is graphically shown in \autoref{F-one-Transformer}(b). 
Residual connection and layer normalization are also applied after each submodule to ensure training stability. The multi-head self-attention mechanism is the core of the model, which evaluates the mutual importance between all tokens in the sequence by calculating the scaled dot product between the query\,($Q$), key\,($K$), and value,($V$), namely, 
\begin{gather}
\text{Attention}(Q,K,V)  = \text{Softmax}\left( \frac{QK^{\up{T}}}{\sqrt{d_K}} \right) V,
\end{gather}
where $d_K$ is the dimension of $K$ vector. 
{This allows every feature token to interact with and incorporate context from all other features in the feature vector.}


After being processed by the $L$ layer encoders, the final output vector corresponding to the [CLS] token is considered as a highly aggregated representation of the entire input feature sequence. Specifically, in our finalized configuration, the model processes an input sequence of $19$ tokens\,(comprising 1 [CLS] token and 18 feature tokens) with an embedding dimension of $d=512$. The encoder stack comprises $L=6$ identical Transformer layers. Within each layer, the data flow preserves the tensor shape of $B \times 19 \times 512$. The Multi-Head Attention module utilizes $h=16$ heads, each with a dimension of $d_\text{head}=32$ (satisfying the projection $16 \times 32 = 512$). The subsequent Feed-Forward Network employs the GEGLU activation function with an expansion factor of 4, expanding the hidden dimension to $2048$ ($4 \times 512$) before projecting back to $512$. Finally, the $512$-dimensional vector corresponding to the [CLS] token is extracted and passed through the output head. This head explicitly consists of a Layer Normalization step followed directly by a single Linear layer ($512 \mapsto 1$), which maps the latent representation to the meson width.

\subsection{Loss function to describe the meson widths}

The loss function serves as the ultimate principle for the neural network model. The core task of the neural network designed before is to precisely predict the total widths of mesons based on the  input quantum numbers. The prediction precision for the $i$th particle can be represented by the relative error $\epsilon_i$ as
\begin{gather}\label{E-epsilon-i}
\epsilon_i = \frac{|\hat y_i -y_i|}{y_i},
\end{gather}
where $\hat y_i$ and $y_i$ denote the predicted value and true value for the $i$th sample, respectively. 
To obtain the globally optimal description, we use the following mean square relative error\,(MSRE), namely,
\begin{gather}
\text{Loss} =\epsilon^2=\frac1n\sum_{i=1}^n \epsilon_i^2= \frac1n\sum_{i=1}^n  \left(\frac{\hat y_i -y_i}{y_i} \right)^2,
\end{gather}
where the summation is over all the training data samples; also we define $\epsilon$ to represent the overall relative error. The loss function ensures that each particle has the same weight in the training.

\subsection{Training and validation}
The division of the dataset into a training and a test set was a physically-motivated compromise choice since only $\sim400$ real mesons can be used to conduct the deep neural network experiment. The training set is composed of the vast majority of conventional $q\bar{q}$ mesons and a subset of relatively well-established exotic candidates. The test set comprises the remaining particles, resulting in an approximate 90\%:10\% ratio. The training set is used for updating model parameters, while the test set remains completely independent throughout training and is used only for final performance evaluation.

Due to the significant range of meson widths\,($10^{-14} \sim 625$\,MeV), direct minimization of the loss function from random initialization leads to numerical instabilities. We therefore employ a two-phase progressive training strategy. In the first $\sim$100 epochs the LogCosh loss function is used for stable initialization. Then we turn to the usual MSRE for the remaining epochs. Early stopping and learning rate scheduling are based on the convergence behavior of the loss function.

In the training implementation, we use the AdamW optimizer\,\cite{Loshchilov2017} to update model parameters with an initial learning rate $\eta=10^{-4}$ and weight 
decay $10^{-4}$. To achieve stable learning rate adjustments, we employ a hybrid strategy of linear warm-up and ReduceLROnPlateau\,\cite{Krizhevsky2012}: in the first 10 epochs, $\eta$ smoothly increases from $10^{-5}$ to the initial value; subsequently, the learning rate decays by a factor of 0.75 when training loss plateaus. The entire training process is conducted with a batch size of 64 for 
a maximum of $\sim$500 epochs, with loss values continuously monitored.

\section{Numerical results of the meson width predictions} \label{sec-4}

\subsection{Overall performance of the model}

The research focuses on the task of describing and predicting the meson widths ranging from $10^{-14}$\,MeV to $625$\,MeV, which poses a serious challenge to any predictive model. In order to systematically evaluate the performance of the model on meson width predictions, we investigate the prediction accuracy from both macro and micro perspectives.

\begin{figure}[h!]
\centering
\includegraphics[width = 1.0\textwidth, angle=0]{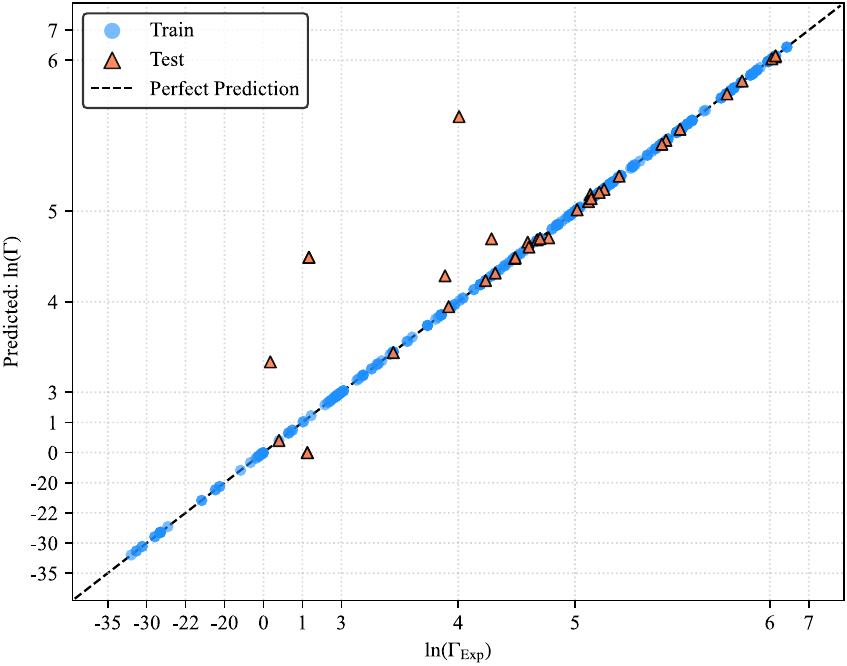}
\caption{The predicted meson widths versus the experimental values in units of MeV under the natural logarithm. The horizontal axis represents the experimental widths reported by PDG, while the vertical axis represents the predicted values. The blue dots denote the results of the training data, while the orange ones denote results of the test sample.} \label{F-actual_vs_predicted}
\end{figure}
In \autoref{F-actual_vs_predicted} we show the overall performance of the meson width predictions for all the data including both the training and the test data. To present the results more clearly across multiple orders of magnitude, we have taken the natural logarithm of the meson widths in units of MeV. The $x$-axis represents the experimental widths reported by PDG\,\cite{PDG2024}, while the vertical $y$-axis represents the predicted values. The blue dots denote the results of the training data while the orange ones denote that of the test sample. It can be seen that the model's predicted values are all closely distributed around the ideal line $y=x$, indicating that the model has successfully captured the overall trend and correlation. To quantitatively evaluate the linear correlation, we calculated the coefficient of determination for the log-transformed values, which is ${R^2=0.9987}$. This macroscopic performance validates the effectiveness of the data processing scheme and the model architecture.

\begin{figure}[h!]
\centering
\includegraphics[width = 1.0\textwidth, angle=0]{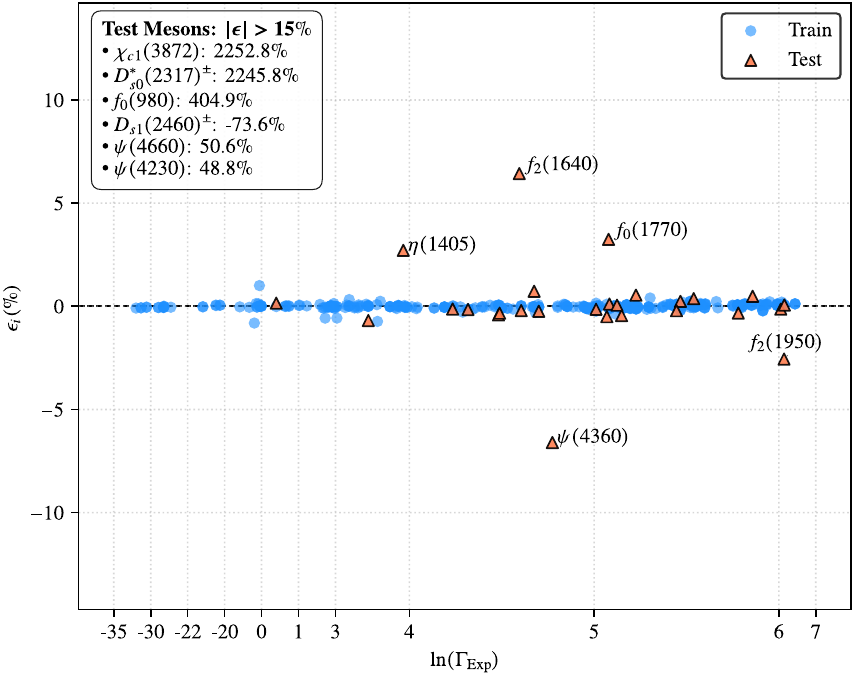}
\caption{The relative error $\epsilon_i$ versus the log widths in units of MeV. The blue\,(orange) denotes the results of the training\,(test) data.} \label{F-width-vs-error}
\end{figure}
To investigate the prediction accuracy at the original physical scale, we also plot the distribution of the relative error $\epsilon_i$\,(defined in \autoref{E-epsilon-i}) versus the log width in units of MeV in \autoref{F-width-vs-error}. The distribution for the training data (blue) is almost clustered around the zero error line.  After excluding several states with $\epsilon_i>48\%$, the obtained relative errors are ${\epsilon=0.12\%, 2.0\%,}$ and {$0.54\%$} for the training set, the test set, and all the data, respectively. More specifically, among the {370} training mesons, $99\%$ of the samples had a relative error ${<1\%}$, while ${80\%}$ of the samples even had a relative error of less than $0.1\%$, which indicates that the model can describe the training data well. 

The generalization ability of the model is more reflected by the prediction performance for the test data. As shown in \autoref{F-width-vs-error}, the error distribution of the test set\,(orange) exhibits high differentiation. On the one hand, most mesons in the test set still show quite low relative errors. After excluding a few outliers with significant errors\,($>48\%$), the mean relative error of the remaining 27 mesons in the test data is about ${2\%}$. If we further exclude the $f$ and $\eta$ families , whose flavor-mixed states require an approximation in the quark content encoding, the remaining {22} traditional mesons have the mean relative errors of about ${0.4\%}$, which achieves the high-precision level comparable to the training set. The above results clearly reveal that the model has good generalization ability in predicting the widths of the traditional mesons.

Notice that there exist significant deviations in the prediction results for those exotic candidates and mixed states, which highlights both the boundary and the utility of our data-driven approach. The prediction deviations are mostly caused by the limited data samples. Statistical analysis reveals that only 21 exotic candidates are included in the training set, which only constitute $\sim 5\%$ of the total samples. Furthermore, a significant portion\,(10/21) of those states have missing quantum numbers. 

As is well known, the performance of the neural network model is basically determined by the volume, distribution and diversity of the data samples. The lack of exotic samples has largely constrained the model's prediction accuracy and generalization ability to the exotic states. As more exotic states are discovered in experiments, we will also incorporate these new particles into the subsequent model updates to improve the performance and generalization.

On the other hand, the large prediction errors for those states, in turn, may indicate their distinct internal structures from traditional mesons. The latest PDG reviews\,\cite{PDG2025_NonQQ,PDG2025_HeavySpec} explicitly discuss that the more favored interpretation for the $\chi_{c1}(3872)$ is a $D\bar{D}^*$ hadronic molecule, while both $D_{s0}^{*}(2317)$ and $D_{s1}(2460)$ are widely considered as $DK$ and $DK^*$ molecular states, respectively. In this context, the deep neural network can serve as a data-driven probe to explore the meson inner structure. A large prediction error may also hint at the particle's inner structure or dynamics deviates from the traditional assumption.
\autoref{tab:theory_comparison} shows a comparison of the model's predictions with theoretical calculations for those test sample mesons. The comparison 
reveals that the relative errors are around $\sim1\%$, 
while theoretical approaches show errors ranging from $\sim8\%$ 
to several hundred percent. Different 
theoretical methods can yield substantially different predictions for the 
same state. For instance, the calculated width for $B_{s2}^*(5840)$ ranges 
from 1.0\,MeV to 10.3\,MeV, corresponding to relative errors of $0.3\%$ and 
$560\%$, respectively.

\begin{table}[h!] 
	\centering
	\caption{Comparison of meson widths\,(in units of MeV) with theoretical predictions  for test sample mesons. $\Gamma_{\text{Exp}}$ denotes experimental width from the PDG\,\cite{PDG2024}; $\Gamma_{\text{Th.1(2)}}$ denotes theoretical predictions from various approaches; $\Gamma_{\text{This}}$ denotes the predictions of this work. Here the relative errors are calculated by Error=$|\Gamma_{\text{pred}} - \Gamma_{\text{Exp}}|/\Gamma_{\text{Exp}} \times 100\%$.}
	\label{tab:theory_comparison}
	\resizebox{\textwidth}{!}{
		\begin{tabular}{ l|c|ccc|ccc|cc }
			\toprule[1.5pt]
			\multicolumn{1}{l|}{Particle} 
			& \multicolumn{1}{c|}{$\Gamma_{\text{PDG}}$} 
			& \multicolumn{1}{c}{$\Gamma_{\text{Th.1}}$} 
			& \multicolumn{1}{c}{Ref.}
			& \multicolumn{1}{c|}{Error (\%)}
			& \multicolumn{1}{c}{$\Gamma_{\text{Th.2}}$}
			& \multicolumn{1}{c}{Ref.}
			& \multicolumn{1}{c|}{Error (\%)}
			& \multicolumn{1}{c}{$\Gamma_{\text{This}}$}
			& \multicolumn{1}{c}{Error (\%)} \\
			\midrule[1.2pt]
			$B_{s2}^*(5840)$ & 1.49 & 1.0 & \cite{PhysRevD.97.114020} & $32.9$ & 10.3 & \cite{PhysRevD.99.094043} & $560$ & 1.49 & 0.2 \\
			\midrule[1.2pt]
			$K_2^*(1430)$ & 100.0 & 80.1 & \cite{2017EPJC...77..861P} & $20.0$ & - & - & - & 99.8 & $0.2$ \\
			\midrule[1.2pt]
			$K^*(1410)$ & 231.8 & 214 & \cite{2017EPJC...77..861P} & $7.7$ & - & - & - & 231.3 & $0.2$ \\
			\midrule[1.2pt]
			$D_1(2420)$ & 31.3 & 25 & \cite{PhysRevD.72.054029} & $20.1$ & - & - & - & 31.1 & $0.7$ \\
			\midrule[1.2pt]
			$D_2(2740)$ & 88.5 & 56.3 & \cite{PhysRevD.88.114003} & $36.3$ & 111.6 & \cite{CPC:10.1088/1674-1137/39/6/063101} & 26.1 & 88.1 & $0.4$ \\
			\midrule[1.2pt]
			$\psi(4360)$ & 118.4 & 37.5 & \cite{2024EPJC...84..810M} & $216$ & 38.6 & \cite{2024EPJC...84..810M} & $207$ & 110.6 & $6.6$ \\
			\bottomrule[1.5pt]
		\end{tabular}
	}
\end{table}

In summary, the analysis in this subsection indicates that the model show high accuracy in predicting the widths of the traditional mesons and good generalization performance. At the same time, it shows systematically significant errors in dealing with the potential exotic states. However, this result in turn provides us a tentative probe to identify and reexamine the potential exotic states from their total widths.

\subsection{Meson width predictions and quantum numbers identification}
We list the meson width spectra obtained by the deep neural network in  the appendix of Supplementary Materials.  
In \autoref{width-prediction}, we provide quantitative width predictions for {15} unmeasured mesons\,(green) in PDG, as well as {32} new mesons\,(orange) that were  theoretically  predicted by the GI model\,\cite{GI1985} but not yet experimentally discovered.  These results provide helpful references and tests for future experimental searches and theoretical studies.

In \autoref{tab:bottomonium_pred} we compare the predictions for the unmeasured  bottomonium mesons  with  the theoretical results from Godfrey-Moats\,(GM) in ref.\,\cite{Godfrey2016}. 
For $\chi_{b1}(nP)$ and $h_b(nP)$, both approaches predict an increasing 
trend with radial quantum number $n$, $\Gamma(1P) < \Gamma(2P)$. For 
$\chi_{b0}(nP)$, our predictions decrease slightly with $n$ while 
GM values remain approximately constant. For $\chi_{b2}(nP)$, 
the trends are opposite, our predictions decrease with $n$ while 
GM predictions increase. Regarding relative magnitudes across 
different $J^{PC}$ states, both approaches predict $\chi_{b0}(nP)$ to be 
broader than $\chi_{b1}(nP)$ and $h_b(nP)$ states at the same radial 
excitation level. The obtained predictions can be tested in the future experiments, which can also serve as a reference for potential experimental explorations. 

Notice that the performance in bottomonium systems is more affected by the limited training data. 
Bottomonium comprises $\sim$2.5\% of our training set, with excited 
$P$-wave and $D$-wave states almost entirely absent. The model transfers 
knowledge from light-quark and charm sectors where training data are 
more abundant. The results also illustrate that the performance of data-driven approaches is more reliable within the training distribution.

\begin{table}[ht] 
	\centering
	\caption{ Comparison of predicted widths\,(MeV) for several unmeasured ($b\bar{b}$) states with theoretical estimates from Godfrey-Moats\,(GM) in ref.\,\cite{Godfrey2016}.}
	\label{tab:bottomonium_pred}
	\resizebox{\textwidth}{!}{
		\begin{tabular}{l|cc|cc|cc|ccc}
			\toprule[1.5pt]
			State & $\chi_{b0}(1P)$ & $\chi_{b0}(2P)$ & $\chi_{b1}(1P)$ & $\chi_{b1}(2P)$ & $h_b(1P)$ & $h_b(2P)$ & $\chi_{b2}(1P)$ & $\chi_{b2}(2P)$ & $\chi_{b2}(3P)$ \\
			$J^{PC}$ & $0^{++}$ & $0^{++}$ & $1^{++}$ & $1^{++}$ & $1^{+-}$ & $1^{+-}$ & $2^{++}$ & $2^{++}$ & $2^{++}$ \\
			\midrule[1.2pt]
			This  & 73.3 & 59.4 & 0.16 & 0.93 & 0.35 & 1.23 & 19.2 & 17.3 & 14.6 \\
			GM & 2.6 & 2.6 & 0.096 & 0.117 & 0.073 & 0.084 & 0.18 & 0.23 & 0.25 \\
			\bottomrule[1.5pt]
		\end{tabular}%
	}
\end{table}
Besides predicting the unknown meson widths, the model also provides us a tentative probe to explore the possible quantum numbers for undetermined mesons based on the comparison between the experimental width and the predicted one. We conducted a series of hypothesis tests on various particles with unknown quantum numbers using the trained model. Namely, the possible quantum numbers are fed into the model to predict the corresponding widths which are then compared with the experimental values.  The comparison results provide the model's suggested quantum numbers for those states, which are listed in \autoref{T-widths-unknow}.
\begin{figure}[t!]
\centering
\includegraphics[width = 1.0\textwidth, angle=0]{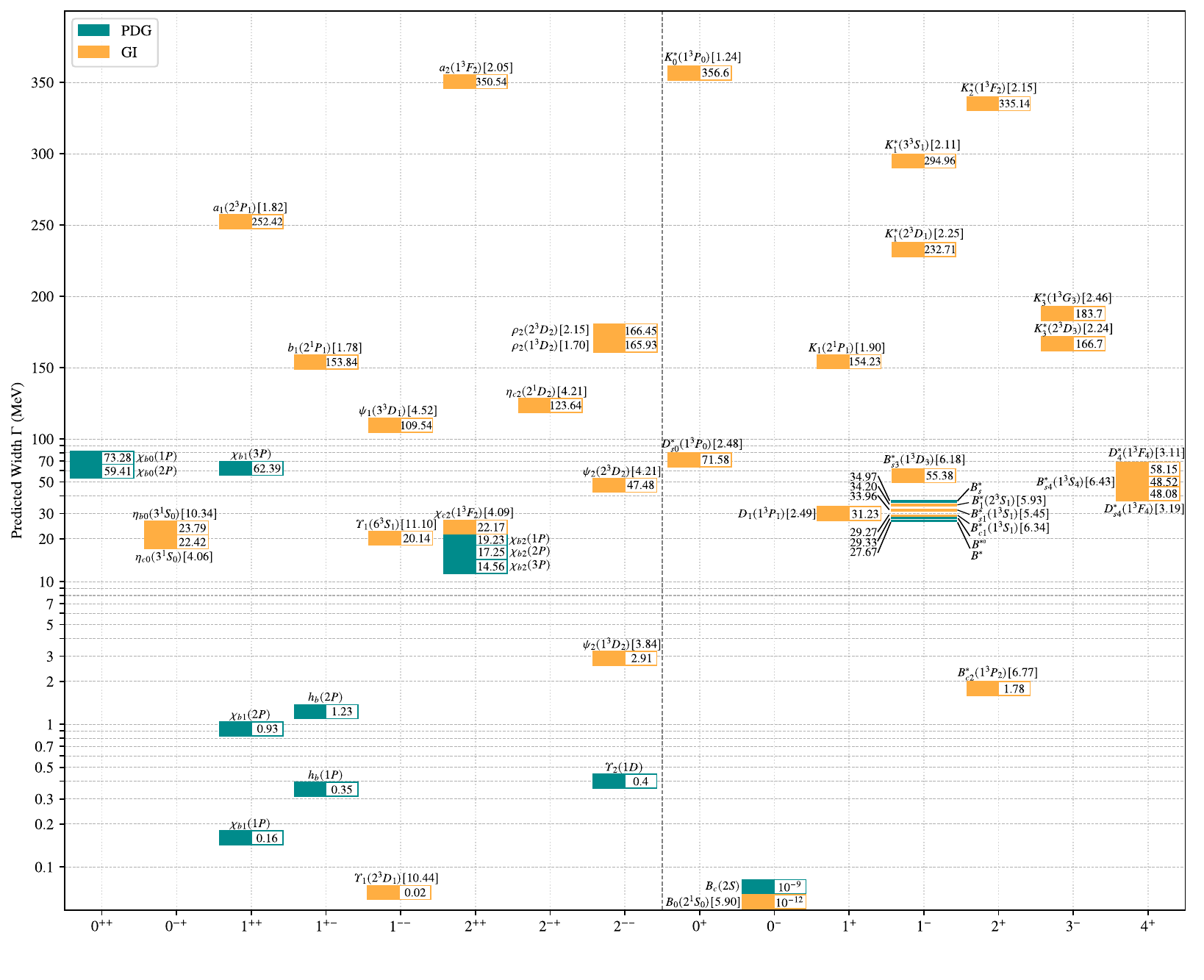}
\caption{The predicted widths  for {47} meson states, where the green labels denote the states observed in experiments and the orange ones denote the states predicted by the GI model\,\cite{GI1985}.} \label{width-prediction}
\end{figure}

\begin{table}[ht]
\caption{The model's predicted widths and favored quantum numbers for the undetermined mesons.
}\label{T-widths-unknow}
\setlength{\tabcolsep}{3pt}%
\vspace{0.2em}\centering
\resizebox{\textwidth}{!}{
\begin{tabular}{ l|cc|ccr|ccccr }
\toprule[1.5pt]
\multicolumn{1}{l|}{States} 	&\multicolumn{1}{c}{$I^{(G)}(J^{P(C)})$} 				&\multicolumn{1}{c|}{$\Gamma_\up{Exp}$}  &\multicolumn{1}{c}{$I^{(G)}(J^{P(C)})$} &\multicolumn{1}{c}{Quark} 			&\multicolumn{1}{c|}{$\Gamma_\up{Pre1}$}	&\multicolumn{1}{c}{$I^{(G)}(J^{P(C)})$}	&\multicolumn{1}{c}{Quark}   &\multicolumn{1}{c}{$\Gamma_\up{Pre2}$}\\
\midrule[1.2pt]
$X(1750)$	&$?^{-}(1^{--})$ 	&$120\pm10$  &$0^{-}(1^{--})$ 	&$s\bar{s}$ &$149.1$ 	&$0^{-}(1^{--})$ 	&$u\bar{u}s\bar{s}$ &$149.1$ \\
\midrule[1.2pt]
$K(1630)$	&$\frac12(?^{?})$ 	&$16^{+19}_{-16}$  &$\frac12(?^{-})$ &$d\bar{s}$ &$16.4$ 	&- 	&- &- \\
$K(3100)$	&$?^{?}(?^{?})$ 	&$42\pm16$  &$\frac12(?^{+})$ &$u\bar{s}$ &$41.5$ 	&$\frac32(2^{+})$ 	&$sd\bar{u}\bar{u}$ &$38.5$ \\
\midrule[1.2pt]
$D^*(2640)$	&$\frac12(?^{?})$ 	&$<15$  &$\frac12(?^{-})$ &$c\bar{d}$ &$16.9$ 	&- 	&- &- \\
$D(3000)$	&$\frac12(?^{?})$ 	&$190\pm80$  &$\frac12(0^{+})$ &$c\bar{u}$ &$186.1$ 	&- 	&- &- \\
$D_{sJ}(3040)$	&$0(?^{?})$ 	&$240\pm60$  &$0(?^{-})$ &$c\bar{s}$ &$177.3$ 	&- 	&- &- \\
\midrule[1.2pt]
$X(3940)$	&$?^{?}(?^{??})$ 	&$43^{+28}_{-18}$  &$0^{-}(3^{+-})$ &$c\bar{c}$ &$36.1$ 	&$0^{+}(2^{++})$ &$c\bar{c}$ &$35.4$ \\
$X(4160)$	&$?^{?}(?^{??})$ 	&$136^{+60}_{-35}$  &$0^{+}(2^{-+})$ &$u\bar{u}c\bar{c}$ &$138.5$ 	&$0^{+}(2^{-+})$ 	&$d\bar{d}c\bar{c}$ &$160.9$ \\
$X(4630)$	&$0^{+}(?^{?+})$ 	&$170^{+140}_{-80}$  &$0^{+}(2^{-+})$ 	&$s\bar{s}c\bar{c}$ &$167.3$&$0^{+}(1^{-+})$ &$s\bar{s}c\bar{c}$ &$152.9$ 	 \\
\midrule[1.2pt]
$B^*_J(5732)$	&$?(?^{?})$ 		&$128\pm18$ 	&$\frac12(?^{-})$ &$d\bar{b}$ &$126.5$ 	&- &- &- \\
$B_J(5840)^+$	&$\frac12(?^{?})$ 		&$220\pm80$ 	&$\frac12(?^{-})$ &$u\bar{b}$ &$190.7$ 	&- &- &- \\
$B_J(5840)^0$	&$\frac12(?^{?})$ 		&$130\pm40$ 	&$\frac12(?^{-})$ &$d\bar{b}$ &$123.9$ 	&- &- &- \\
$B_J(5970)^+$	&$\frac12(?^{?})$ 		&$62\pm20$ 	&$\frac12(?^{-})$ &$u\bar{b}$ &$74.5$ 	&- &- &- \\
$B_J(5970)^0$	&$\frac12(?^{?})$ 		&$81\pm12$ 	&$\frac12(?^{-})$ &$d\bar{b}$ &$90.5$ 	&- &- &- \\
\midrule[1.2pt]
$B_{sJ}^*(5850)$	&$?(?^{?})$ 	&$47\pm22$ 	&$0(1^{-})$ &$s\bar{b}$ &$36.9$ 	&- &- &- \\
$B_{sJ}(6063)$	&$0(?^{?})$ 		&$26\pm6$ 	&$0(2^{-})$ &$s\bar{b}$ &$22.9$ 	&- &- &- \\
$B_{sJ}(6114)$	&$0(?^{?})$ 		&$66\pm28$ 	&$0(4^{-})$ &$s\bar{b}$ &$82.7$ 	&- &- &- \\
\midrule[1.2pt]
${{ T}_{{{c {{\bar{ c}}}}}}{(4050)}}$	&$1^{-}(?^{?+})$ 	&$82^{+50}_{-28}$  &$1^{-}(3^{-+})$ &$u\bar{d}c\bar{c}$ &$82.8$&$1^{-}(2^{-+})$ &$u\bar{d}c\bar{c}$ &$105.0$ 	 \\
${{ T}_{{{c {{\bar{ c}}}}}}{(4055)}}$	&$1^{+}(?^{?-})$ 	&$45\pm13$  &$1^{+}(3^{+-})$ &$u\bar{d}c\bar{c}$ &$49.1$&$1^{+}(2^{+-})$ &$u\bar{d}c\bar{c}$ &$35.9$ 	 \\
${{ T}_{{{c {{\bar{ c}}}}}}{(4100)}}$	&$1^{-}(?^{?+})$ 	&$150^{+80}_{-70}$  &$1^{-}(1^{-+})$ &$u\bar{d}c\bar{c}$ &$164.3$&$1^{-}(0^{++})$ &$u\bar{d}c\bar{c}$ &$230.4$ 	 \\
${{ T}_{{{c {{\bar{ c}}}}}}{(4250)}}$	&$1^{-}(?^{?+})$ 	&$180^{+320}_{-70}$  &$1^{-}(1^{-+})$ &$u\bar{d}c\bar{c}$ &$175.2$&$1^{-}(0^{-+})$ &$u\bar{d}c\bar{c}$ &$172.0$ 	 \\
${{ T}_{{{c c {{\bar{ c}}} {{\bar{ c}}}}}}{(6600)}}$	&$0^{+}(?^{??})$ 	&$440^{+230}_{-200}$\cite{PhysRevLett.132.111901}  &$0^{+}(2^{++})$ &$cc\bar{c}\bar{c}$ &$96.2$&$0^{+}(1^{+-})$ &$cc\bar{c}\bar{c}$ &$120.3$\\
${{ T}_{{{c c {{\bar{ c}}} {{\bar{ c}}}}}}{(6900)}}$	&$0^{+}(?^{??})$ 	&$137\pm21$  &$0^{+}(2^{++})$ &$cc\bar{c}\bar{c}$ &$99.3$&$0^{+}(1^{+-})$ &$cc\bar{c}\bar{c}$ &$122.4$\\
${{ T}_{{{c c {{\bar{ c}}} {{\bar{ c}}}}}}{(7100)}}$	&$0^{+}(?^{??})$ 	&$97^{+40}_{-29}$\cite{PhysRevLett.132.111901}  &$0^{+}(2^{++})$ &$cc\bar{c}\bar{c}$ &$100.6$&$0^{+}(0^{++})$ &$cc\bar{c}\bar{c}$ &$80.4$\\
${{ T}_{{{b {{\overline{ s}}}}}}{(5568)}}$	&$1(?^{?})$ 	&$19^{+9}_{-7}$  &$1(3^{+})$ &$u\bar{d}s\bar{b}$ &$23.2$&$1(1^{+}/2^{+})$ &$u\bar{d}s\bar{b}$ &$25.7$ 	 \\

\bottomrule[1.5pt]
\end{tabular}
}
\end{table}

For the ${X(4160)}$ state, the PDG favors the $J^{PC}=2^{-+}$ assignment with a significance of over 4$\sigma$. Under the assumption of a $2^{-+}$ $[c\bar{c}u\bar{u}]$ tetraquark state, the predicted width 138.5\,MeV is  well consistent with the experimental value 136\,MeV. For ${X(4630)}$, the PDG favors the $1^{-}$ assignment over the $2^{-}$; the model shows that under the $[c\bar{c}s\bar{s}]$ hypothesis, both the $1^{-+}$ and $2^{-+}$ yield predictions consistent with the experimental width. For ${T_{c\bar{c}}(4100)}$, the PDG suggests the assignment $0^{+}$ or $1^{-}$. The model's prediction $164.3$\,MeV under the $1^{-+}$ hypothesis is quite close to the experimental value $150^{+80}_{-70}$\,MeV and then favors the latter assignment. The model shows good consistency between its predictions and the preliminary conclusions drawn from experimental analyses by the PDG, which in turn provides independent, data-driven support for these findings.

For particles where the PDG has not provided a clear preference, the model offers independent assignments for their quantum numbers. For example, the $J^{P}=0^{+}$ hypothesis for ${D(3000)}$ yields the predicted width with about a 2\% relative error from the experimental value. For the tetraquark ${T_{c\bar{c}}(4050)}$,  under the $3^{-+}$ assignment the model gives a predicted width that is quite close to the experimental value. Plausible quantum number assignments were also identified for several other particles, such as ${X(1750)}$ and ${B_{sJ}^{*}(5850)}$.

For the $K$, $D$ and $B$ states,  if only the parity $P$ is specified, the predictions show good consistency with the experimental data, which is a manifestation of the strategy for the missing features. In these cases, our analysis provides support for the particles to have specific $P$ values, while the determination of their precise spin $J$ values is beyond the scope of the current predictive model.

\subsection{Analysis of quark content encoding}
\begin{figure}[t!]
	\centering
	\includegraphics[width = 1.0\textwidth, angle=0]{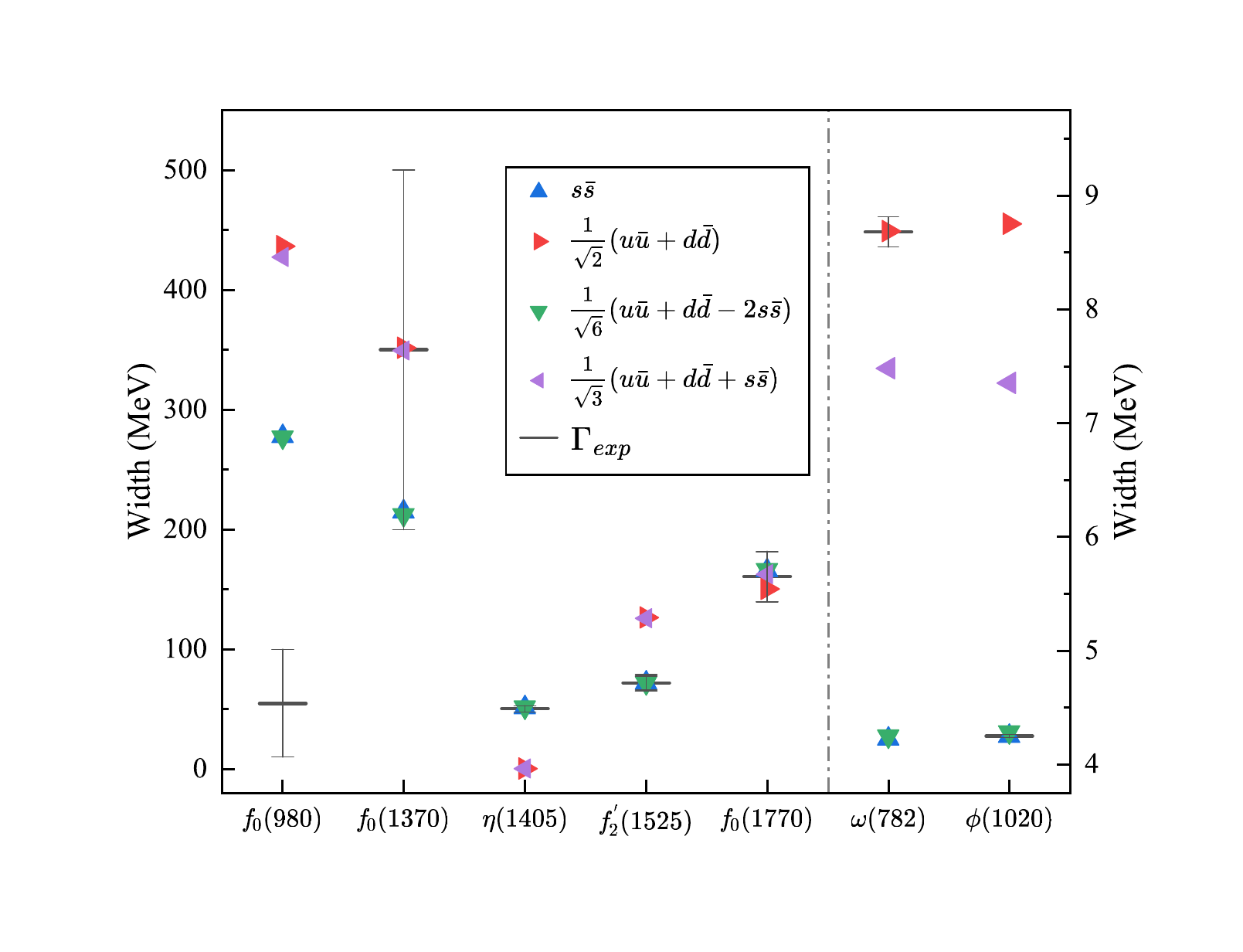}
	\caption{Analysis of the predicted widths for representative mesons under different quark content hypotheses. The black horizontal lines indicate the experimental central values taken from the PDG\,\cite{PDG2024}. The colored markers represent the model predictions corresponding to four specific encoding assumptions: pure light quarks $[\frac{1}{\sqrt2}(u\bar{u}+d\bar{d})$, $\triangleright$], pure strange quarks $[s\bar{s}$, $\triangle$], singlet-like mixing $[\frac{1}{\sqrt3}(u\bar{u}+d\bar{d}+s\bar{s})$, $\triangleleft$], and octet-like mixing [$\frac1{\sqrt6}(u\bar{u}+d\bar{d}-2s\bar{s})$, $\triangledown$]. The double $Y$-axis is used to accommodate the large range of widths, with the right axis specifically for the narrow states $\omega(782)$ and $\phi(1020)$.}
	\label{F-sensitivity}	
\end{figure}

To evaluate the influence of encoding approximations and explore the internal structures of mixed states, we performed a sensitivity analysis on seven representative mesons. We tested the model's response to four physical hypotheses: $\frac{1}{\sqrt2}(u\bar{u}+d\bar{d})$,  ($s\bar{s}$), $\frac1{\sqrt6}(u\bar{u}+d\bar{d}-2s\bar{s})$, and $\frac{1}{\sqrt3}(u\bar{u}+d\bar{d}+s\bar{s})$. The results are summarized in \autoref{F-sensitivity}. A general pattern emerges from this analysis: the predictions tend to cluster according to the dominant quark flavor. Specifically, the results for the $s\bar{s}$ and octet-mixing scenarios\,(both strange-rich) are numerically close. This suggests that the model is sensitive to the relative weight of strangeness in the input.

We first examined this sensitivity using well-established states. According to the PDG review\,\cite{PDG2025_QuarkModel}, the $\phi(1020)$ is characterized as a nearly pure $s\bar{s}$ state. Similarly, for $f_2'(1525)$, the review cites a mixing angle of $\alpha_T \approx 81^\circ$, which is proximal to the ideal mixing limit of $90^\circ$\,(corresponding to a pure $s\bar{s}$ configuration). This indicates that the $f_2'(1525)$ is predominantly composed of strange quarks. The model's predictions indicate that for both $\phi(1020)$ and $f_2'(1525)$, the relative errors are minimal\,($<1\%$) only under the strange-rich encoding\,($s\bar{s}$ or octet), whereas the light-quark hypotheses lead to significant deviations. Conversely, for the $\omega(782)$, the pure light $\frac1{\sqrt2}(u\bar{u}+d\bar{d})$ encoding yields the optimal prediction, suggesting that the model correctly captures the correspondence between flavor dynamics and total width. 

Regarding the $\eta(1405)$, predictions based on the light-quark structure  show a significant deviation from the experimental width. In contrast, the strange-rich scenarios align much better with the data, with the pure $s\bar{s}$ hypothesis yielding a relative error of $\sim2\%$. This numerical preference for strangeness is interesting in light of the discussion in the PDG review on the $\eta(1405/1475)$ region\,\cite{PDG2025_LightResonances}. While the review suggests that the nearby $\eta(1475)$ is the likely $s\bar{s}$ member of the nonet due to its $K^*\bar{K}$ decay, it also explicitly indicates that the data are not sufficient to exclude a sizeable $s\bar{s}$ admixture in the $\eta(1405)$ as well. The model's output is consistent with the presence of such a strange quark component. 

For the scalar $f_0(980)$, we observe that all tested $q\bar{q}$ encoding hypotheses yield relative errors exceeding $400\%$. These large deviations indicate that the current feature representation, based on standard quark-antiquark assignments, is insufficient to describe this state. This observation aligns with the view summarized in the PDG review, which notes a consensus towards a predominantly four-quark nature for the $f_0(980)$\,\cite{PDG2025_ScalarMesons}.

Lastly, for the $f_0(1770)$, the model's response is less discriminatory compared to other states, with relative errors remaining below $7\%$ across all four hypotheses. The flavor-singlet encoding yields the lowest error $\sim1.4\%$. This state may have a more complex nature as documented in the PDG review, where its interpretation involves varying mixtures of $q\bar{q}$ and gluonic components\,\cite{PDG2025_LightResonances}.

\subsection{Symmetries reproduced by the neural network}

A successful model should not only have high accuracy in predicting numerical values, but it should also comply with fundamental physical principles and symmetries. In this subsection we aim to verify whether the deep neural network model can autonomously understand and reproduce the fundamental symmetries in particle physics, for example, the charge conjugation symmetry and the approximate isospin symmetry of meson widths.

\subsubsection{Charge conjugation symmetry}

The $CPT$ theorem is one of the deepest results of quantum field theory, which states that the combined operation of time reversal, charge conjugation, and parity in any order is an exact symmetry of any interaction.  An important experimental implication of the $CPT$ theorem claims that every particle must have precisely  the same mass and lifetime\,(width) as its antiparticle, which has been tested by the $K^0$-$\bar K^0$ mass difference fraction being less than $6\times10^{-19}$\,\cite{PDG2024}. In the above research, the particle and antiparticle are trained by the same neural network model and the only difference is the feature vector of the data set. In meson feature vector $\bm{v}$, a particle and its antiparticle are directly different by a charge conjugation transformation to the quark contents, for example, 
\begin{gather}
\bm{q}_{B_c^+}=(0,0,0,0,0,0,1,0,0,1) ~\rightleftharpoons~(0,0,0,0,0,0,0,1,1,0) =\bm{q}_{B_c^-}.
\end{gather}
If the trained model can also predict the same width for a particle and its antiparticle, we can deduce that the neural network model is invariant under the charge conjugation transformation. 

\begin{figure}[h!]
\centering
\includegraphics[width = 1.0\textwidth, angle=0]{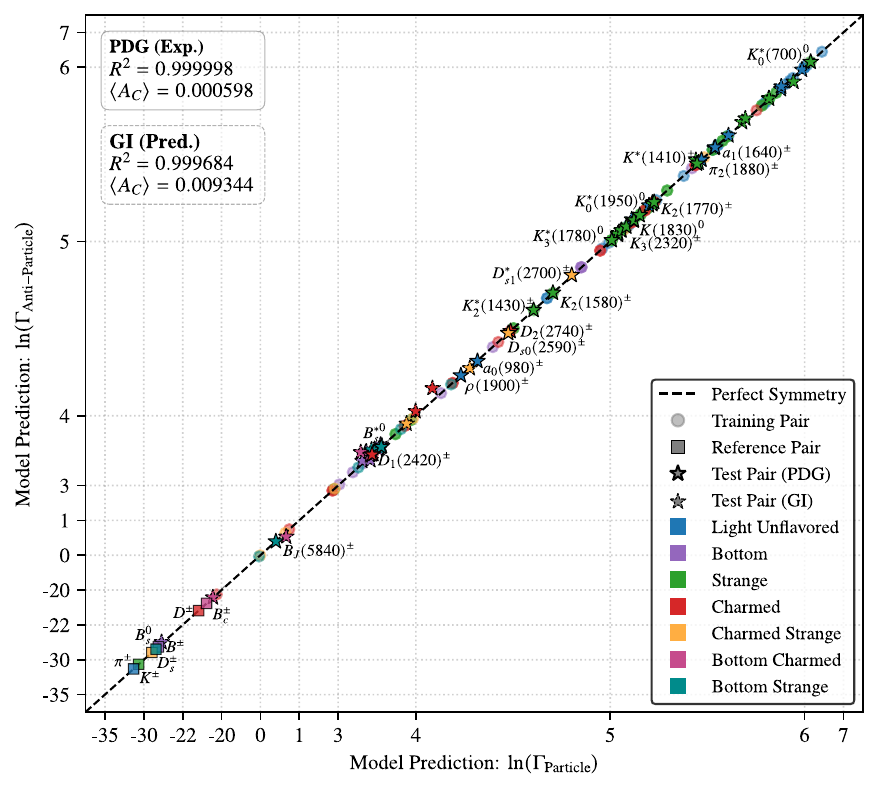}
\caption{The predicted widths for particle-antiparticle pairs, plotted on a logarithmic scale. The points in the test set are represented by pentagrams, while those in the training set are represented by circles. Key reference particles within the training set are highlighted with squares. The theoretically predicted (GI) pairs are marked with dashed edges to demonstrate generalization to unseen states. Different colors are used to denote mesons with different quark contents.} \label{F-C-symmetry}
\end{figure}

In \autoref{F-C-symmetry}, we show the invariance of charge conjugation transformation clearly. The $x$-axis denotes the predicted log widths for particles, while the $y$-axis represents the  log widths for the corresponding antiparticles. Then each point in this plot represents a two-dimensional coordinate $(\ln \Gamma_\text{Particle}, \ln \Gamma_\text{Antiparticle})$ .  If the model was really invariant under the transformation for particle and antiparticle, all the points should be located closely along the line $y=x$.  From  \autoref{F-C-symmetry}, we can see that all the data points, regardless of their quark composition (color distinction) or role in the dataset (shape distinction), are strictly listed along the $y=x$ line. We also calculate the determination coefficient to be ${R^2=0.9999}$. In order to reflect the degree of this symmetry violation for the neural network model, we introduce the violation coefficient as
\begin{gather}
A_{Ci} = \frac{\Gamma_i -\Gamma_{\bar i}}{\Gamma_i + \Gamma_{\bar i} },
\end{gather}
where $\Gamma_i$ and $\Gamma_{\bar i}$ denote the widths of the particle $i$ and its antiparticle, respectively. Then we can define the root mean square violation coefficient $A_C$ to denote the overall violation of charge conjugation symmetry of the model for all mesons, 
\begin{gather}
A_C = \sqrt{\braket{A^2_{Ci}}} =\sqrt{\frac1n\sum A_{Ci}^2} ={0.06\%}.
\end{gather}
The obtained result indicates that the model makes almost identical width predictions for particles and their antiparticles. To further validate the charge conjugation symmetry on the unseen data, we also presented the behaviors for charge conjugation symmetry on undiscovered particles in \autoref{F-C-symmetry}, where the undiscovered particles are indicated by the dashed markers and the meson masses are based on the theoretical predictions of the famous GI model\,\cite{GI1985}. The model still maintains a strong symmetry with a violation coefficient of $A_C = 0.9\%$ and a determination coefficient of $R^2 = 0.9997$, demonstrating robust generalization beyond the training distribution. From above analysis, it can be deduced that the trained neural network model is invariant under the charge conjugation transformation.

\subsubsection{Isospin symmetry}

Within the dynamics of the Standard Model, the strong interaction behaves much larger than the electromagnetic interaction and the weak interaction. Meson widths are then usually determined by the corresponding strong decay behaviors except for those that can only electroweakly decay. On the other hand, the strong interaction is approximately invariant under the isospin transformation, and then the widths of mesons that can undergo strong decay also exhibit approximate isospin symmetry. In this subsection we investigate  whether the deep neural network model can capture this deep abstract symmetry. 
 
\begin{figure}[h!]
\centering
\includegraphics[width = 1.0\textwidth, angle=0]{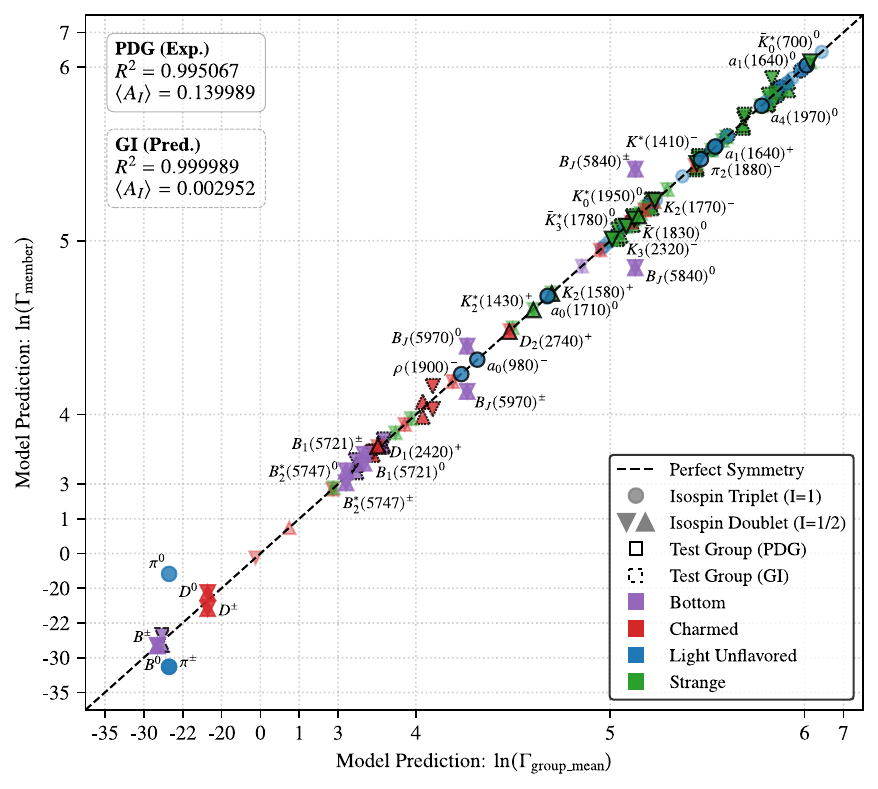}
\caption{The predicted width  versus the averaged width of the isospin multiplets for mesons with isospin $I>0$. The $x$-axis denotes the log average widths of the isospin multiplets, while $y$-axis denotes the predicted width of a single state. The dots represent results for the isospin triplets $I=1$, while the triangles represent those for the isospin doublets $I=1/2$.The predictions for the unseen GI multiplets are indicated by markers with dashed edges.} \label{F-I-symmetry}
\end{figure}

In order to check this point, we plot the predicted widths versus the averaged widths for those isospin multiplets in \autoref{F-I-symmetry}. The results show the approximate isospin symmetry of the model by comparing the predicted meson width\,($y$-axis) with the average one of its isospin multiplets\,($x$-axis).
As shown in \autoref{F-I-symmetry}, the data points predicted by the model are also clustered along the $y=x$ line which stands for the perfect isospin symmetry of meson widths. The corresponding determination coefficient of this isospin symmetry are determined to be ${R^2_I=0.9950}$. One can see that the model correctly identifies the isospin multiplets and maintain good consistence within their internal members. The behaviors of the test samples marked with black boxes in the figure validate the good generalization performance of the model.

The violation coefficient $A_{Ii}$ for isospin symmetry can be introduced similarly as
\begin{gather}
A_{Ii} = \frac{\Gamma_{i} - \braket{\Gamma_{i}}_I}{ \Gamma_{i} + \braket{\Gamma_{i}}_I },
\end{gather}
where $\Gamma_i$ denotes the width for meson $i$, and $\braket{\Gamma_i}_I$ denotes  the average one of the isospin multiplets. After calculating the root mean square for $A_{Ii}$, we obtain the violation coefficient of isospin symmetry for the model as
\begin{gather}
A_I = \sqrt{\braket{A_{Ii}^2}} = \sqrt{\frac1n \sum A_{Ii}^2 } = {14.0\%},
\end{gather}
where the summation is over all the mesons of isospin $I>0$. This value  reveals that the isospin symmetry is significantly violated. {A closer inspection reveals that the deviation is not uniformly distributed across all multiplets. The pion multiplet ($\pi^{\pm}, \pi^0$) is the predominant source of this large value. If excluding the pions, the violation coefficient for the remaining multiplets drops to 5.0\%. A less pronounced effect is observed for the $D$ meson doublets ($D^{\pm}, D^0/\bar{D}^0$); removing them as well reduces the coefficient to a remarkable 3.0\%. This final value is mainly a reflection of the more subtle deviations within the $B$ mesons sector.} Notice the isoscalar mesons are not included in the $A_I$ calculation. It can be seen clearly from above results that the model still shows reasonable  isospin symmetry but this symmetry is obviously and much more violated compared with the charge conjugation symmetry.

We also verified the isospin symmetry behaviors on the unseen data of the GI predictions, which are indicated by the dashed markers in \autoref{F-I-symmetry} . These theoretical states exhibit a consistent pattern with a violation coefficient of $A_I = 1.4\%$ ($R^2 = 0.9979$), confirming that the model's approximate isospin symmetry generalizes well to the unobserved meson spectrum. These results show that the data-driven symmetries have emerged in the neural network and generalized to the unknown region.

In fact, the model has also captured the violation effects for different symmetries during the training process, and hence reproduced the effects in the inference process.  For the mesons in the training set, the  outputs of the model reflect the patterns it has learned from the input data. For example, the model provides  almost the same widths for $\pi^+$ and $\pi^-$ but quite different width values for $\pi^0$ and $\pi^+$; for $K$, $D$, and $B$ family mesons, the predicted widths behave slightly different within the isospin multiplets but behave quite close for the particles and their corresponding antiparticles. 
 
The difference in the determination coefficients $R^2$\,({0.9999\,vs.\,0.9950}) can be taken as a measure of the strictness of model symmetries. The result is consistent with  physical reality since the charge conjugation symmetry is much more precise than the isospin symmetry. Combining the results in \autoref{F-C-symmetry} and \autoref{F-I-symmetry}, we can conclude that the output of the model not only qualitatively follows the physical symmetry, but also quite precisely reflects the accuracy of different symmetries at a quantitative level. The results above indicate that the constructed feature system and neural network have good learning and inference abilities to distinguish the detailed physics determined by different interactions or input parameters.

The results above indicate that the constructed neural network has good learning and inference abilities to distinguish the complicated physics determined by different interactions and input parameters. It is worth noting that the landscape of machine learning in QCD is quite diverse. The symbolic regression has been demonstrated demonstrate effective in discovering analytical expressions for fundamental parameters\,\cite{Wang2024CPL}, while physics-informed networks are powerful in solving governing equations for dynamical systems such as flux tubes\,\cite{Kou2025PRD}. In parallel, our work explores the ability of deep neural networks in predicting hadron widths where explicit analytical formulations remain challenging. Collectively, these distinct data-driven approaches contribute to a  neural-network-based understanding of the strong interaction physics.

\section{Summary and outlook}

In this work, we build a deep neural network model to predict the total widths of mesons based on the Transformer architecture, which can also be used as a tentative probe to determine the meson quantum numbers. The input feature vector is constructed from the meson quantum numbers and mass. To overcome the lack of data samples for training the deep learning model, we adopt the Gaussian Monte-Carlo data enhancement method to enhance the meson data by considering the experimental errors, which also significantly improves the robustness and generalization performance of the model. Within the width range from $\sim10^{-14}$ to $\sim625$\,MeV,  the relative errors of the model's predictions are ${\epsilon=0.12\%, ~2.0\%,}$ and {$0.54\%$} in the training set, the test set, and all the meson data, respectively. We present width predictions for currently discovered mesons and some theoretical predicted ones. The obtained width spectra can provide useful references to both the experimental and theoretical research and can also be tested by the future experiments. Its predictive deviations further serve as a valuable probe to study the quantum numbers and inner structures for some undetermined mesons. 

Furthermore, the obtained model exhibits well-defined charge conjugation symmetry and the approximate isospin symmetry consistent with real physical phenomena. Namely, the pure data-driven model can spontaneously reproduce symmetries in the real physical world. The results indicate that the deep neural network can serve as an independent supplementary  research paradigm to describe and explore the hadron structures and the complicated  interactions in particle physics besides traditional experimental measurements, theoretical calculations, and lattice simulations.

Based on the current study and above discussions, we can see that there 
exist several potential optimization directions for this data-driven 
neural network model. First, a more precise encoding scheme for the quark 
content would be important for the exotic states and mesons with flavor 
mixing, such as the $\eta$, $\phi$, $f$ family, and the unnatural open flavor 
mesons. For those states, the experimental good quantum numbers cannot 
uniquely determine the mesons. A more effective training method can be 
explored to maximize the utilization of existing experimental data. Also, 
a more efficient encoding scheme can be explored to decrease the number of 
parameters and to eliminate the possible encoding redundancy of the current 
model.

\acknowledgments
This work is supported by the National Key R\&D Program of China\,(2022YFA1604803), and the Natural Science Basic Research Program of Shaanxi\,(2025JC-YBMS-020). It is also supported by  the National Natural Science Foundation of China\,(NSFC) under Grant Nos.\,12047503, 12575097, 12005169, 12075301, and 11821505.

\medskip

\bibliographystyle{JHEP}
\setlength{\bibsep}{1ex}  

\providecommand{\href}[2]{#2}\begingroup\raggedright\endgroup

\appendix
\section{Spectra of the predicted meson widths}\label{sec-0}

\begin{figure}[h!]
\centering
\includegraphics[width = 0.95\textwidth, angle=0]{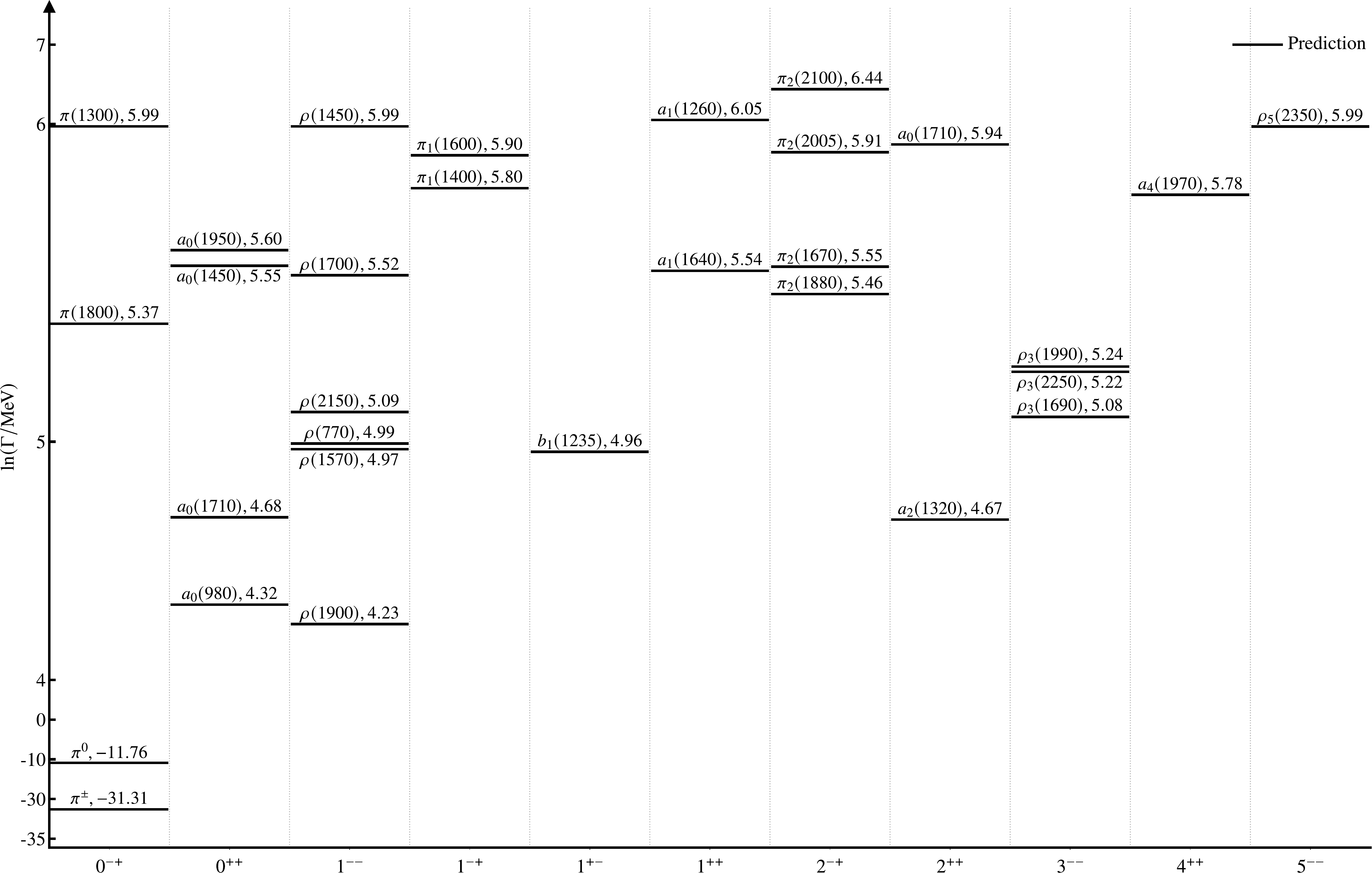}
\caption{The predicted log widths  for mesons with quark contents $[u\bar d, \sqrt{1/2}(u\bar u-d\bar d), d\bar u]$. The $y$-axis denotes the log widths $\ln \Gamma$ with $\Gamma$ in units of MeV. } \label{F-width-ud}
\end{figure}

\begin{figure}[h!]
\centering
\includegraphics[width = 0.95\textwidth, angle=0]{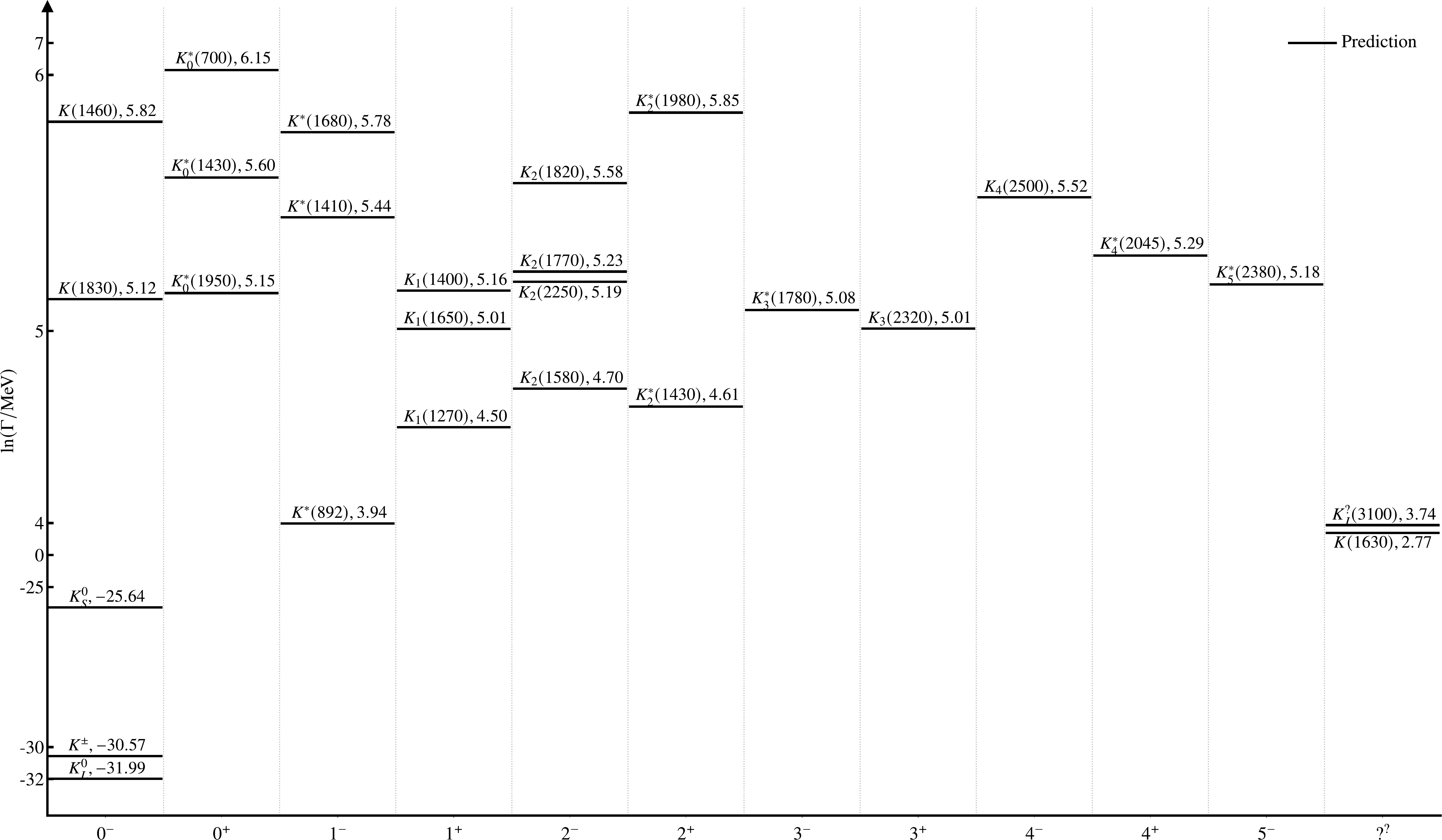}
\caption{The predicted log widths  for strange mesons $(u\bar s,d\bar s)$. The legend is as for \autoref{F-width-ud}.} \label{F-width-usds}
\end{figure}

\begin{figure}[h!]
\centering
\includegraphics[width = 0.88\textwidth, angle=0]{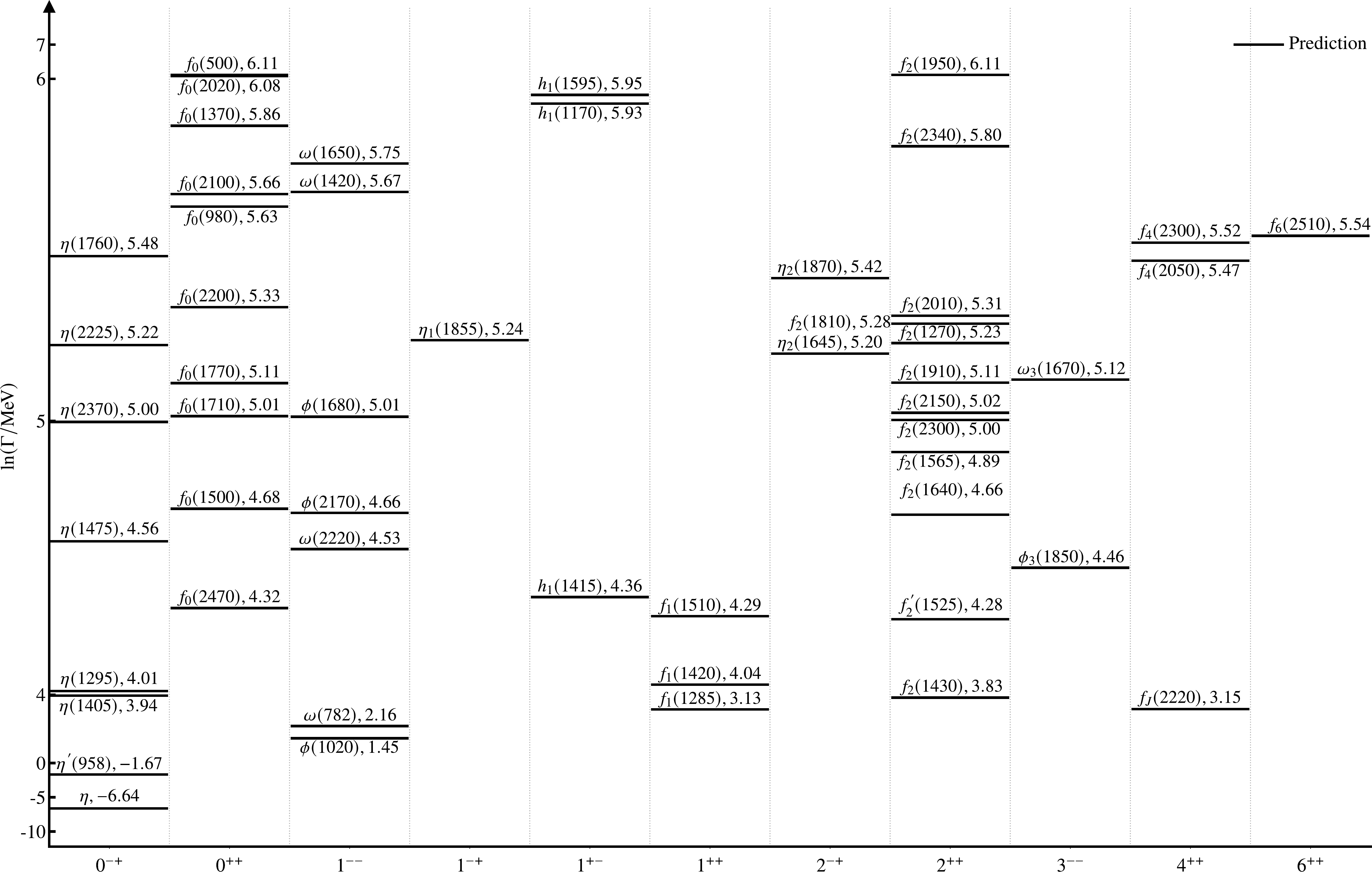}
\caption{The predicted log widths  for isoscalar mesons $(u\bar u,d\bar d, s\bar s)$. The legend is as for \autoref{F-width-ud}.} \label{F-width-uuddss}
\end{figure}

\begin{figure}[h!]
\centering
\includegraphics[width = .88\textwidth, angle=0]{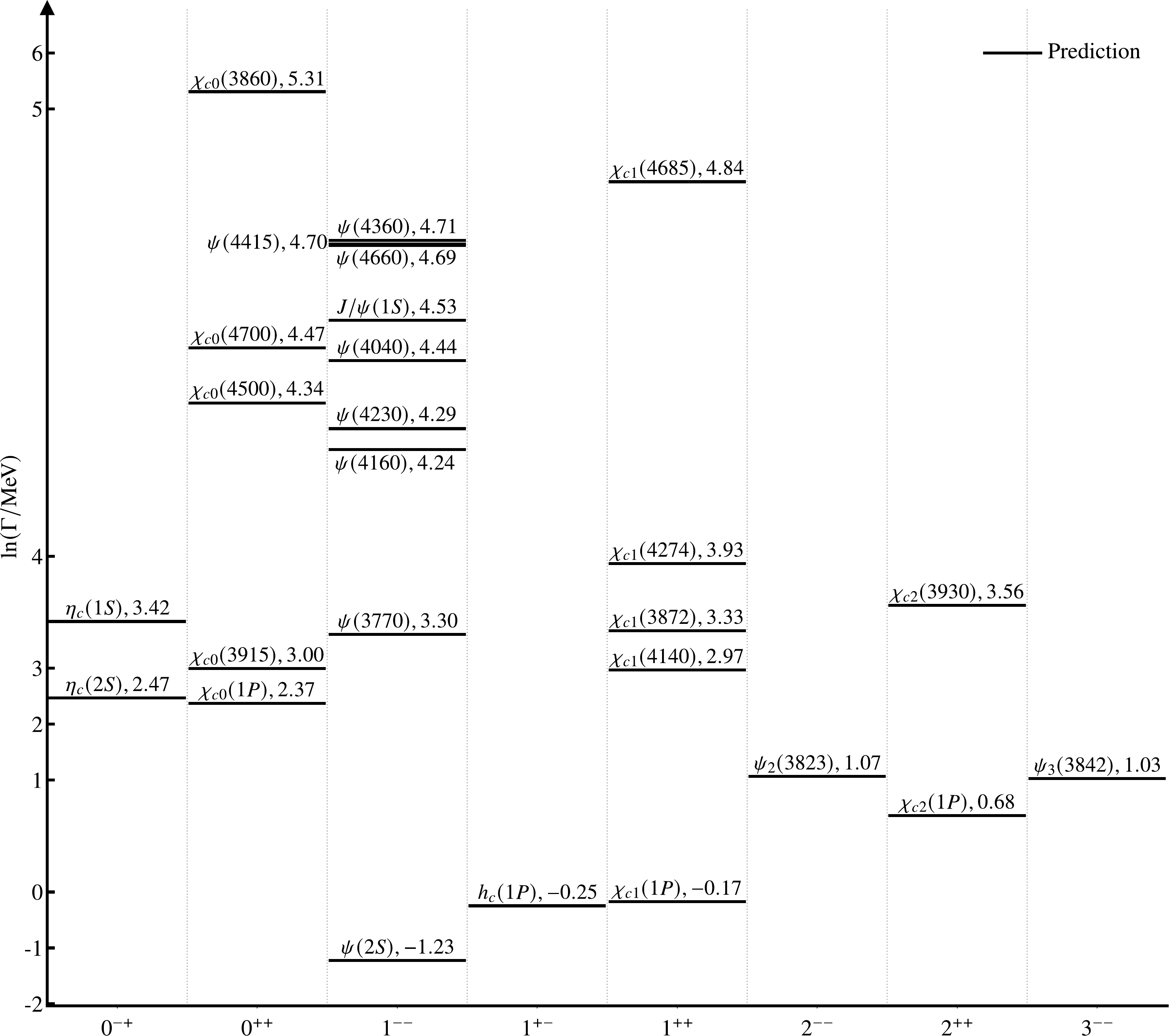}
\caption{The predicted log widths  for charmonia mesons $(c\bar c)$. The legend is as for \autoref{F-width-ud}.} \label{F-width-cc}
\end{figure}

\begin{figure}[h!]
\centering
\includegraphics[width = 1.0\textwidth, angle=0]{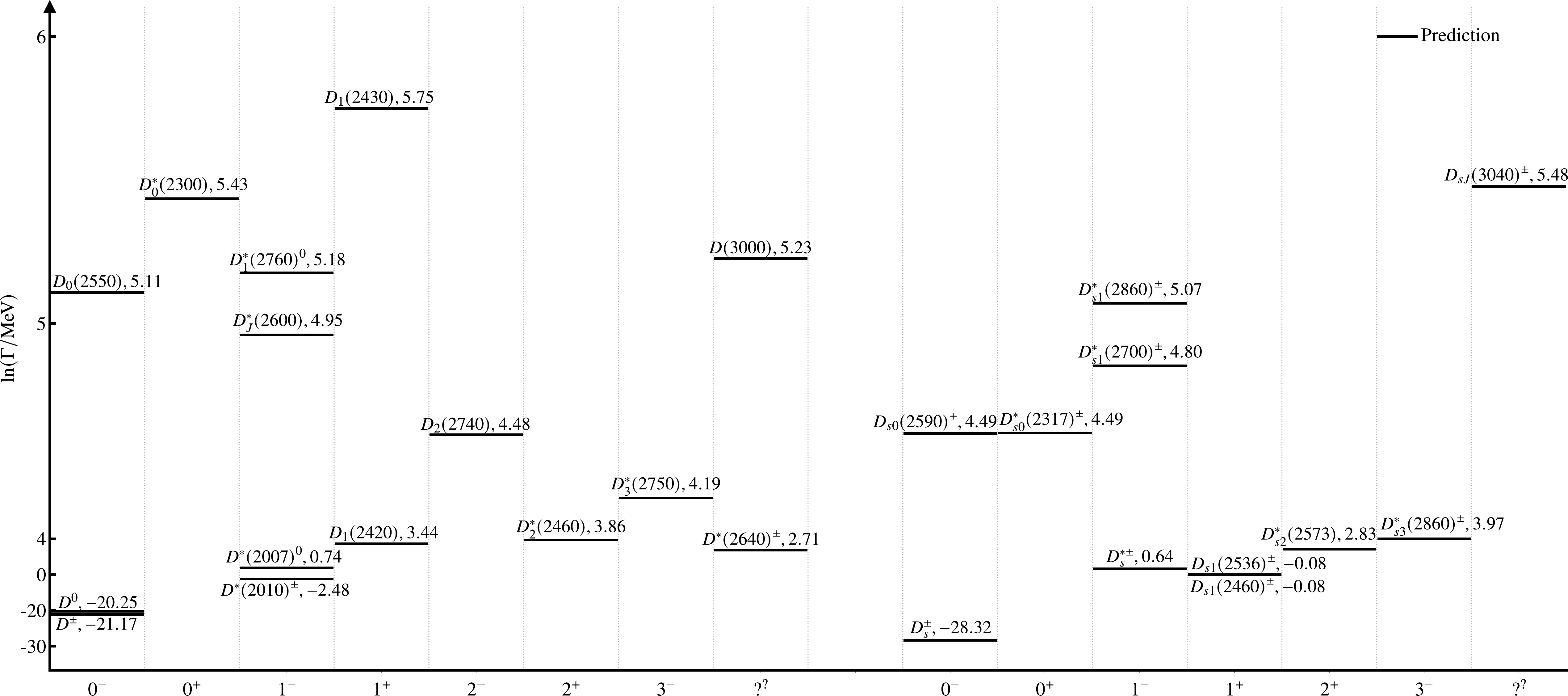}
\caption{The predicted log widths  for charmed mesons $(c\bar d, c\bar u, c\bar s)$ with the legend as for \autoref{F-width-ud}.} \label{F-width-cdcucs}
\end{figure}

\begin{figure}[h!]
\centering
\includegraphics[width = 1.0\textwidth, angle=0]{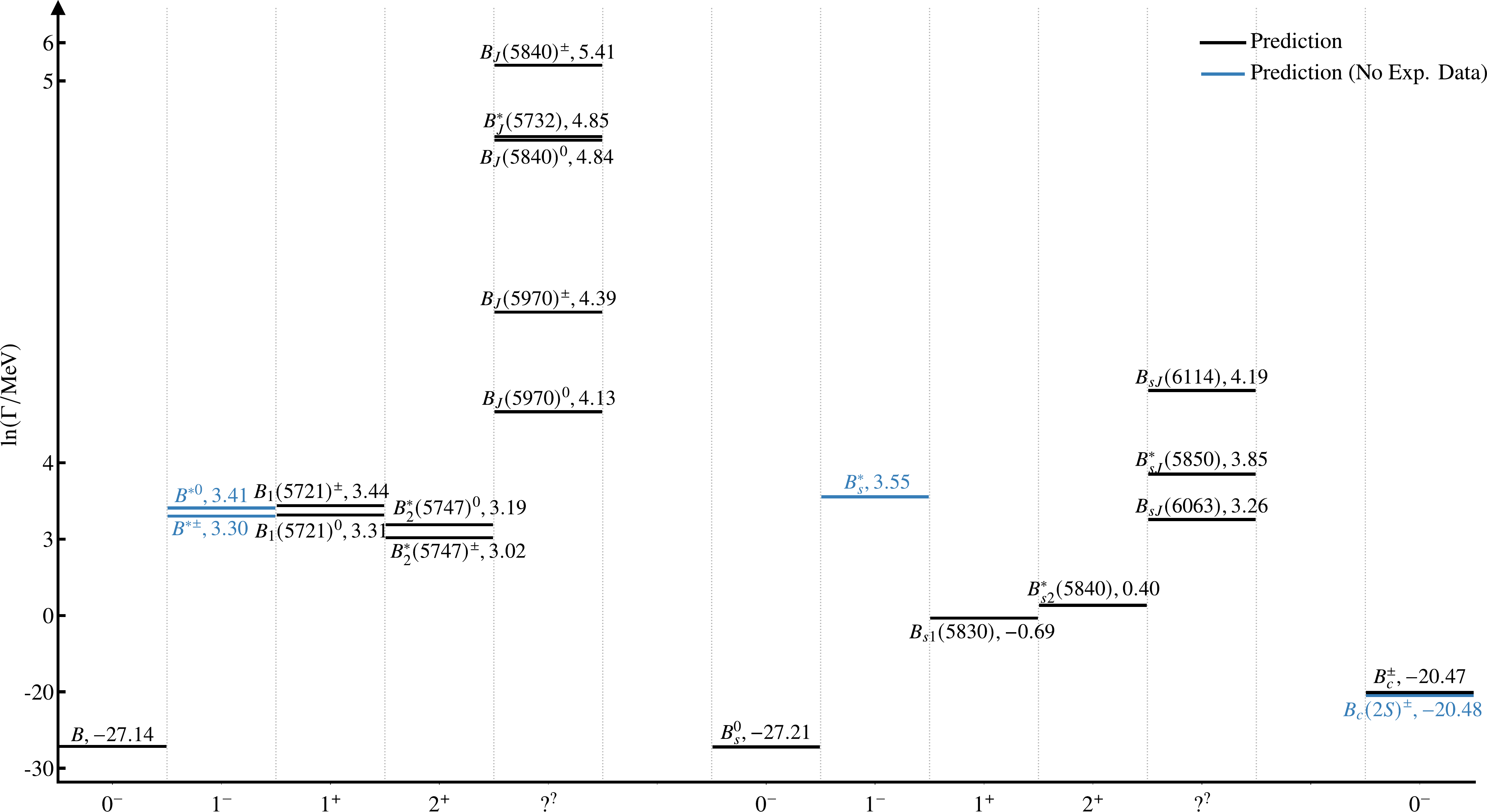}
\caption{The predicted log widths  for bottomed mesons $(b\bar u, b\bar d, b\bar s, b\bar c)$. The legend is as for \autoref{F-width-ud}.} \label{F-width-bubdbsbc}
\end{figure}

\begin{figure}[t]
\centering
\includegraphics[width = 0.95\textwidth, angle=0]{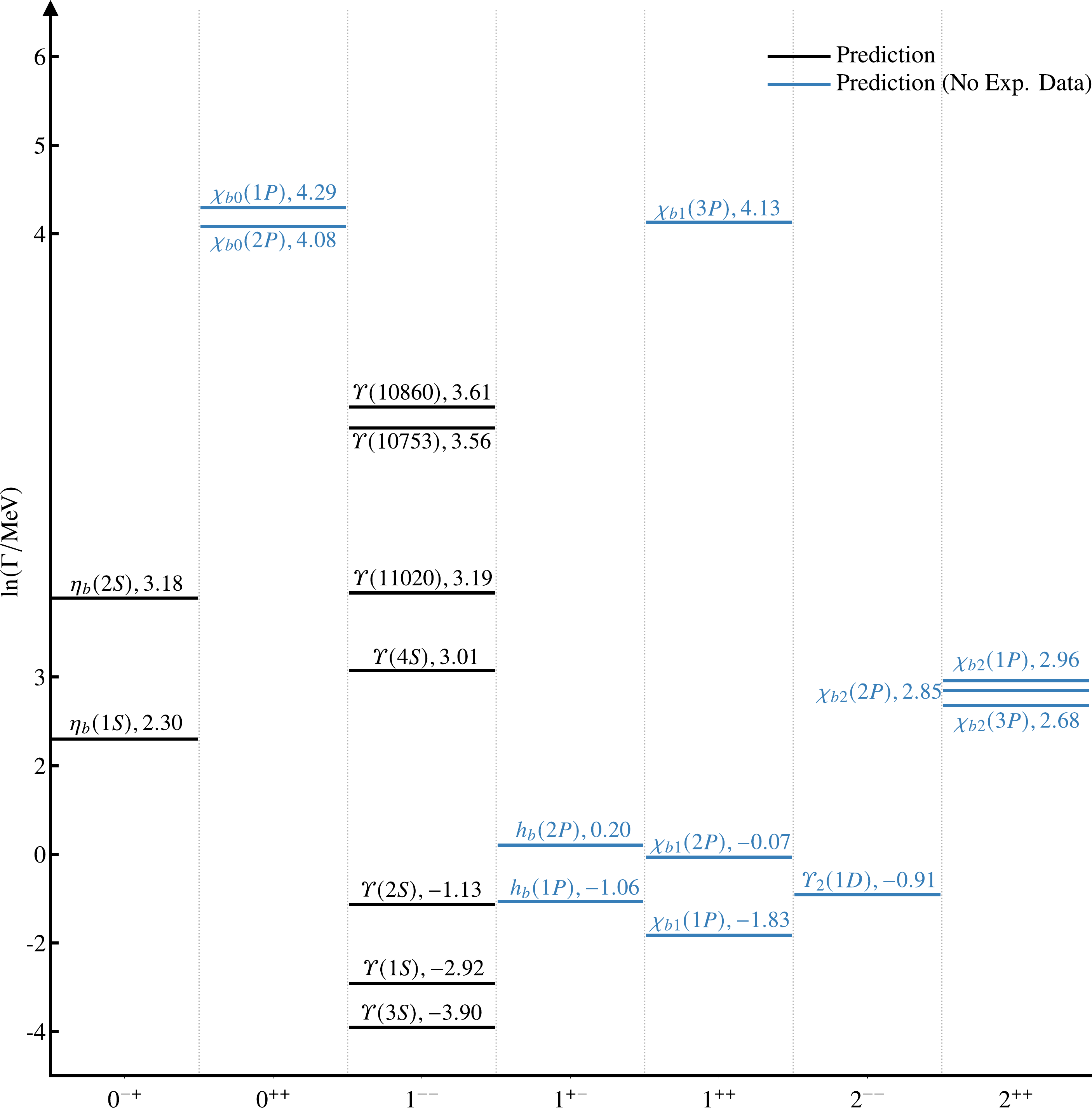}
\caption{The predicted log widths  for $b\bar b$ mesons. The legend is as for \autoref{F-width-ud}.} \label{F-width-bb}
\end{figure}

We list the spectra of the predicted meson widths in \autoref{F-width-ud} $\sim$ \autoref{F-width-bb}.
\autoref{F-width-ud} shows the predicted widths for the isovector mesons  with quark contents $[u\bar d, \sqrt{1/2}(u\bar u-d\bar d), d\bar u]$; \autoref{F-width-usds}, the strange mesons $(u\bar s,d\bar s)$; \autoref{F-width-uuddss}, the isoscalar mesons, $(u\bar u,d\bar u, s\bar s)$; \autoref{F-width-cc}, the charmonia mesons $(c\bar c)$; \autoref{F-width-cdcucs}, the charmed mesons $(c\bar d, c\bar u, c\bar s)$; \autoref{F-width-bubdbsbc}, the bottomed mesons $(b\bar u, b\bar d, b\bar s, b\bar c)$; and finally, \autoref{F-width-bb}, the $\Upsilon$ family $(b\bar b)$.

\end{document}